\documentclass[twocolumn,prd,showpacs,preprintnumbers,amsmath,amssymb,superscriptaddress,floatfix,altaffilletter]{revtex4-1}
\usepackage[dvips]{graphicx,color}
\usepackage{array,hhline,dcolumn}
\usepackage[pagewise,mathlines]{lineno}
\usepackage{rotating}
\usepackage{epsfig}
\usepackage{subfigure}
\usepackage{amsmath}
\usepackage{amsfonts}
\usepackage{amssymb}
\usepackage{longtable}
\usepackage{graphicx}%

\begin{document}

\title{Improved Parameterization of K$^{+}$ Production in p-Be Collisions at Low Energy\\Using Feynman Scaling}
\date{\today}
\author{C. Mariani}\affiliation{Columbia University, New York, NY 10027}
\author{G. Cheng}\affiliation{Columbia University, New York, NY 10027}
\author{J. M. Conrad}\affiliation{Department of Physics, Massachusetts Institute of Technology, Cambridge, Massachusetts 02139, USA}
\author{M. H. Shaevitz}\affiliation{Columbia University, New York, NY 10027}

\begin{abstract}
This paper describes an improved parameterization for proton-beryllium production of
secondary $K^{+}$ mesons for experiments with primary proton beams from 8.89 to
24~GeV/c. The parameterization is based on Feynman scaling in which the
invariant cross section is described as a function of $x_{F}$ and $p_{T}$.
This method is theoretically motivated and provides a better description of
the energy dependence of kaon production at low beam energies than
other parameterizations such as the commonly used \textquotedblleft Modified Sanford-Wang\textquotedblright\  model.
This Feynman scaling parameterization has been used for the
simulation of the neutrino flux from the Booster Neutrino Beam at Fermilab and has been
shown to agree with the neutrino interaction data from the SciBooNE experiment.
This parameterization will also be
useful for future neutrino experiments with low primary beam energies, such as
those planned for the Project X accelerator.

\end{abstract}


\maketitle


\section{Introduction}

This paper describes a parameterization for inclusive production of
secondary $K^{+}$ mesons in proton-beryllium collisions,
\begin{equation}
p+Be\rightarrow K^{+}+X.
\end{equation}
for experiments with low primary proton beam energies ranging in kinetic energy from below 9
to 24~GeV. The parameterization is based on Feynman scaling(F-S)~\cite{FeynmanPRL},
in which the invariant cross section is described as a function of
transverse momentum, $p_{T}$, and a scaling variable,
$x_{F}={p_{\shortparallel}^{CM}}/{p_{\shortparallel}^{CM~max}}$.
Various scaling parameterizations are known to describe data
well above $\sim$20~GeV~\cite{Bonesini:2001iz,Norbury:2009}. In
this paper, we show that the F-S form describes data down to 8.89~GeV/c beam momentum. This result
provides an alternative model to the traditional \textquotedblleft Modified
Sanford-Wang\textquotedblright\ \cite{SanWang,Wang:1970} parameterization used to describe
secondary production at low primary proton beam momentum. The results from this
F-S analysis have been used in the neutrino flux parameterization
of the Booster Neutrino Beam (BNB) at Fermilab and has been checked against measurements by the SciBooNE experiment~\cite{SciBooNE}.
This parameterization will be useful for future neutrino experiments using low primary proton beam energies.

\begin{figure}[hptb!]
\begin{center}
\includegraphics[width = 0.8\columnwidth]{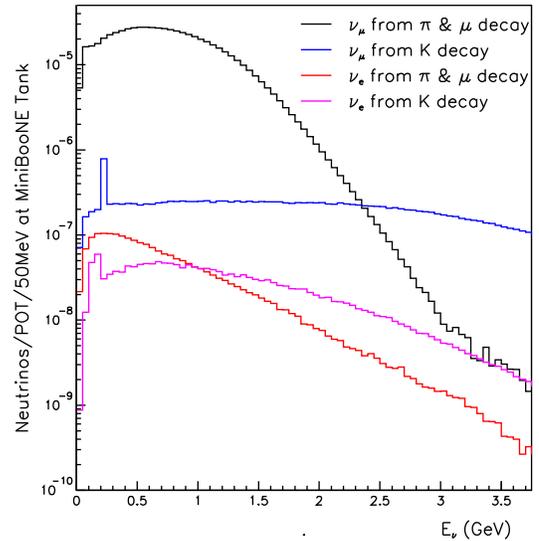}
\caption{Predicted $\nu_{\mu}$ and $\nu_{e}$ flux spectrum from decaying
pions, kaons, and muons for the BNB and SciBooNE and MiniBooNE experiments.}%
\label{fig:pr_flux}%
\end{center}
\end{figure}

The primary motivation for this work was the simulation of neutrinos in the
BNB line. This line provides neutrinos for the
MiniBooNE~\cite{MiniBooNE} and SciBooNE~\cite{SciBooNE} experiments, as well
as possible future experiments, including the upcoming MicroBooNE ~\cite{MicroBooNE}
experiment. In this beam line, protons with 8~GeV kinetic energy are directed onto a 1.8
interaction length beryllium target. The charged pions and kaons which are
produced are focused by a magnetic horn into a 50~m decay region, where they
subsequently decay to produce neutrinos. The average energy of $\pi^{+}$
($K^{+}$) that decay to neutrinos in the MiniBooNE detector acceptance is 1.89
(2.66)~GeV. Therefore 37.6\% (92.1\%) of the particles decay before the end of the 50~m long
decay region. The most relevant decay modes for MiniBooNE are
$\pi^{+}\rightarrow \mu^{+}\nu_{\mu}$, $K^{+}\rightarrow\mu^{+}\nu_{\mu}$,
which produce 99.4\% of the neutrino beam, and $K^{+}\rightarrow\pi^{0}%
e^{+}\nu_{e}$, $\mu^{+}\rightarrow e^{+}\bar{\nu}_{\mu}\nu_{e}$,
$K_{L}^{0}\rightarrow\pi^{-}e^{+}\nu_{e}$, and $K_{L}^{0}\rightarrow\pi
^{+}e^{-}\bar{\nu}_{e}$, which produce the remaining 0.6\%.

Figure~\ref{fig:pr_flux} shows the predicted flux for the BNB line at the
MiniBooNE detector. While the flux is predominately due to $\pi^{+}$ decay, the
$K^{+}$ decay is the dominant source above 2~GeV. The $\nu_e$ flux from kaon decay
contributes one of the important backgrounds for neutrino oscillation searches
looking for $\nu_e$ appearance.
In addition, the kaon neutrino flux provides
an interesting source of high energy events for experiments on the
BNB line for studying neutrino cross sections.
Therefore, it is important for the BNB line experiments to have a good
first-principles prediction of $K^{+}$ production.

A first-principles prediction for $K^{+}$ production is obtained from fitting
data from secondary production experiments with primary beam momentum ranging
from 8.89 to 24~GeV/c. Nine data sets are considered, but only seven are used in the fit as it will be
explained in Section~\ref{sec:fit_machinery}. Because these data are taken at a
range of beam energies, the data must be fit to a parameterization including
changes with beam momentum in order to scale the result to the 8.89~GeV/c of
the BNB line momentum.

\subsection{Feynman Scaling Formalism}\label{sec:FSformalism}

Over the past several decades, many experiments have made measurements of
particle production by protons of various energies on many different nuclear
targets. These data have been used to study the phenomenology of particle
production and have led to several scaling laws and quark counting rules. For
inclusive particle production, Feynman put forward a theoretical
model~\cite{FeynmanPRL} where
the invariant cross section is only a function of $x_{F}$
 and $p_{T}$. The
invariant cross section is related to the
commonly used differential cross section by:

\begin{eqnarray}
\frac{d^{2} \sigma}{dpd\Omega}=\frac{p^{2}}{E}E\frac{d^{3}\sigma}{dp^{3}}.
\end{eqnarray}

Defining

\begin{eqnarray}
E \frac{d^{3}\sigma}{dp^{3}} = A F(x_{F},p_{T}),
\end{eqnarray}

this leads to:

\begin{eqnarray}
\frac{d^{2} \sigma}{dpd\Omega}=\frac{p^{2}}{E}A\ F(x_{F},p_{T}).
\label{eq:d2dpdOmega}
\end{eqnarray}

\noindent
$A$ is a factor and $F$ is the F-S function that depends on $x_F$ and $p_T$.
The quantity $p_{\shortparallel}^{CM\text{ }\max}$, which appears in the
denominator of the definition of $x_F$,  depends upon the particle being
produced and is derived from the exclusive channels given in Table~\ref{ExclusiveChannels}.

\begin{table}[h!] \centering
\begin{tabular}
[c]{ccccc}\hline
Produced & Exclusive & M$_{X}$ & $\sqrt{s_{thresh}}$ & E$_{thresh}^{beam}$\\
Hadron & Reaction & (GeV/c$^{2})$ & (GeV) & GeV\\\hline
$\pi^{+}$ & pn$\pi^{+}$ & 1.878 & 2.018 & 1.233\\
$\pi^{-}$ & pp$\pi^{+}\pi^{-}$ & 2.016 & 2.156 & 1.54\\
$\pi^{0}$ & pp$\pi^{0}$ & 1.876 & 2.011 & 1.218\\
$K^{+}$ & $\Lambda^{0}pK^{+}$ & 2.053 & 2.547 & 2.52\\
$K^{-}$ & pp$K^{+}K^{-}$ & 2.37 & 2.864 & 3.434\\
$K^{0}$ & p$\Sigma^{+}K^{0}$ & 2.13 & 2.628 & 2.743\\\hline
\end{tabular}
\caption{Threshold production channels for proton + proton production of various mesons.
The exclusive reaction is the final state with the minimum mass, $M_X$.
$\sqrt{s_{thresh}}$ and E$_{thresh}^{BEAM}$ are the threshold center of mass (CM) and laboratory energy.}
\label{ExclusiveChannels}%
\end{table}%

Feynman scaling has been demonstrated for secondary meson production at
primary beam energies above
$\sim$15 to 20~GeV \cite{feynmanhiE,Bonesini:2001iz,Norbury:2009}; this paper
demonstrates the validity of F-S at lower primary beam energies for $K^{+}$ production.
One might expect F-S to be a
better parameterization of $K^{+}$ production than the
\textquotedblleft Modified Sanford-Wang\textquotedblright\  formalism
for two reasons. First, the F-S parameterization properly
accounts for the kinematic effects of the large kaon mass
where even at $x_{F}=0$, the outgoing kaon can have a significant laboratory momentum.
Second, the functional form of the parameterization typically has peak production at
$x_{F}=0$. This is in contrast to the
\textquotedblleft Modified Sanford-Wang\textquotedblright\ formalism, where the
production rate continues to grow as $x_{F}$ becomes more negative.

\subsection{Feynman Scaling Parameterization for the Particle Production Cross
Section}

The Feynman model can be used to describe the expected $x_{F}$ and $p_{T}$
dependence using theoretically inspired functions for these dependences. For
the $x_{F}$ dependence, a parameterization proportional to $\exp(-a\left\vert
x_{F}\right\vert ^{b})$ or $\left(  1-\left\vert x_{F}\right\vert \right)
^{c}$ has the properties consistent with a flat rapidity plateau around
$x_{F}=0$. The expectation of a limited $p_{T}$ range is provided by including
exponential moderating factors for powers of $p_{T}$.

Using this guidance, a F-S parameterization has been developed to
describe kaon production. In order to allow some coupling between the $x_{F}$
and $p_{T}$ distribution an additional exponential factor has been added that
uses the product, $\left\vert p_{T}\times x_{F}\right\vert $.
The $c_i$'s are the seven coefficients of the F-S function.
The kinematic threshold constraint for $K^{+}$ production is imposed by setting $\dfrac
{d^{2}\sigma}{dpd\Omega}$ equal to zero for $\left\vert x_{F}\right\vert >1$.

Including these factors, the final parameterization has the form:

\begin{eqnarray}
&& \frac{d^{2}\sigma}{dpd\Omega} = \frac{p_{K}^{2}}{E_{K}}\left( E_{K}\frac{d^{3}\sigma}{dp_{K}^{3}}\right)=   \nonumber
\left(  \frac{p_{K}^{2}}{E_{K}}\right)c_{1} \times
\\ &&\times \exp\left[c_{3}\left\vert x_{F}\right\vert^{c_{4}}-c_{7}\left\vert p_{T}\times x_{F}\right\vert ^{c_{6}}-c_{2}p_{T}-c_{5}p_{T}^{2}\right]
\label{Scaling_Function}%
\end{eqnarray}


\subsection{The \textquotedblleft Modified
Sanford-Wang\textquotedblright\  Parameterization}\label{sec:SWformalism}

Many neutrino experiments have used the \textquotedblleft Modified
Sanford-Wang\textquotedblright\ parameterization\cite{SanWang,Wang:1970} (S-W):

\begin{eqnarray}
&& \frac{d^{2}\sigma}{dpd\Omega} = c_{1}{p_{K}}^{c_{2}}(1-\frac{p_{K}}{p_{BEAM}-c_{9}}) \times \nonumber
\\ && \times \exp\left[ {{\frac{-c_{3}{p_{K}}^{c_{4}}}{{p_{BEAM}}^{c_{5}}}-c_{6}\theta_{K}(p_{K}-c_{7}p_{BEAM}cos^{c_{8}}\theta_{K})}}\right]
\label{SW_Function}
\end{eqnarray}

This functional form allows for some phenomenological
parameterization of the variations associated with beam energy and process
thresholds. As noted in one of the initial Sanford-Wang papers \cite{SanWang,Wang:1970},
the coefficients for $\pi^{+}$ production are approximately given by:
$c_{2}=0.5,c_{4}=c_{5}=1.67,$ and the $\cos\theta$ term is negligible. With
these substitutions, the formula shows a close although not perfect
relationship with F-S (see Eq.~\ref{Scaling_Function})%

\begin{eqnarray}
E\frac{d^{3}\sigma}{dp^{3}}^{Sanford-Wang}=A^{\prime}\ F^{\prime}(X)\text{
}e^{-Cp_{T}},
\end{eqnarray}

where

\begin{eqnarray}
F^{\prime}(X)=X^{1/2}(1-X)e^{-BX^{5/3}}%
\end{eqnarray}

and

\begin{eqnarray}
X=\frac{p}{p_{BEAM}}%
\end{eqnarray}

Therefore, the S-W fits to the K$^{+}$ data will show only
approximate consistency with F-S.
At low beam energy, produced particle mass effects can become important. Table~\ref{ExclusiveChannels}
gives the minimum mass channels, their invariant mass,
and the beam energy threshold for different particle production processes. In
the S-W formula, the parameter $c_{9}$ is included to approximately provide
the kinematic limit for produced particle momentum. Investigations of the
exact kinematic threshold for $K^{+}$ production show that the maximum $p_{K}$
is approximately equal to $P_{BEAM}-P_{Diff}$ where $P_{Diff}$ varies from 1.7
to 2.2~GeV as $\theta_{K}$ goes from 0 to 0.3~rad. One would therefore
expect that $c_{9}$ would take on values similar to $P_{Diff}$. On the other
hand, the factor $(1-\frac{p_{K}}{p_{BEAM}-c_{9}})$ introduces violations of
the scaling behavior away from this limiting region.

\begin{table*}[htbp!]
\begin{center}
\begin{ruledtabular}
\begin{tabular}
[c]{l|l|c|c|c|c|c|c}
{  K}$^{+}${Data} & {Ref.} & $P_{B}${(GeV/c)}
& {  P}$_{K}${(GeV/c)} & $\theta_{K}${  (rad)} &
${  x}_{F}$ & $p_{T}${(GeV/c)} & ${  \sigma}_{Norm}$\\\hline
{  Abbott} & \cite{abbott} & {  14.6} & ${  2-8}$ & ${  0.35}{  -0.52}$ & ${  -0.12-0.07}$ & ${  0.2-0.7}$ & {  10\%}\\
{  Aleshin} & \cite{aleshin} & {  9.5} & ${  3-6.5}$ & ${  0.06}$ & ${  0.3-0.8}$ & ${  0.2-0.4}$ & {  10\%}\\
{  Allaby} & \cite{allaby} & {  19.2} & ${  3-16}$ & ${  0}{  -0.12}$ & ${  0.3-0.9}$ & ${  0.1-1.0}$ & {  15\%}\\
{  Dekkers} & \cite{dekkers} & {  18.8 , 23.1} & ${  4-12}$ & ${  0}{  ,0.09}$ & ${  0.1-0.5}$ & ${  0.0-1.2}$ & {  20\%}\\
{  Eichten} & \cite{eichten} & {  24.0} & ${  4-18}$ & ${  0}{  -0.10}$ & ${  0.1-0.8}$ & ${  0.1-1.2}$ & {  20\%}\\
{  Lundy} & \cite{lundy} & {  13.4} & ${  3-6}$ & ${  0.03}{  ,0.07}{  ,0.14}$ & ${  0.1-0.6}$ & ${  0.1-1.2}$ & {  20\%}\\
{  Marmer} & \cite{marmer} & {  12.3} & ${  0.5-1}$ & ${  0}{  ,0.09}{  ,0.17}$ & ${  -0.2--0.05}$ & ${  0.0-0.15}$ & {  20\%}\\
{  Piroue} & \cite{piroue} & {  2.74} & ${  0.5-1}$ & ${  0.23}{  ,0.52}$ & ${  -0.3-1.0}$ & ${  0.15-0.5}$ & {  20\%}\\
{  Vorontsov} & \cite{vorontsov} & {  10.1} & ${  1-4.5}$ & ${  0.06}$ & ${  0.03-0.5}$ & ${  0.1-0.25}$ & {  25\%}\\
\end{tabular}
\caption{Data sets for $K^+$ production with proton momentum lower 24 GeV/c. $P_{B}$ indicates the beam momentum and $\sigma_n$ gives the normalization error for the experimental data.}
\label{Data Sets}%
\end{ruledtabular}
\end{center}
\end{table*}

An additional problem with the S-W parameterization is that most of the
function parameters ($c_i$) will be effectively fixed by the scaling
constraints, and this will be limiting the flexibility of the function to match the
$x_{F}$ and $p_{T}$ behavior. The parameter $c_{2}$, for example, should be
close to unity to provide the conversion from invariant to differential cross
section. The parameter $c_{9}$ needs to be approximately equal to 2.0~GeV to
provide the maximum $p_{K}$ dependence, and the parameters $c_{4}$ and $c_{5}$
should be equal in order to preserve a basic $x_{F}$ dependence. Thus, the
S-W parameterization has very little flexibility to fit the data
distributions over the full kinematic range and therefore a formalism like Feynman
scaling is required. In many of the following plots, we will compare prediction results coming
from S-W and F-S parameterizations.

\section{External Data Sets and Kinematic Coverage}

\begin{figure}[hptb!]
\includegraphics[width=0.8\columnwidth]{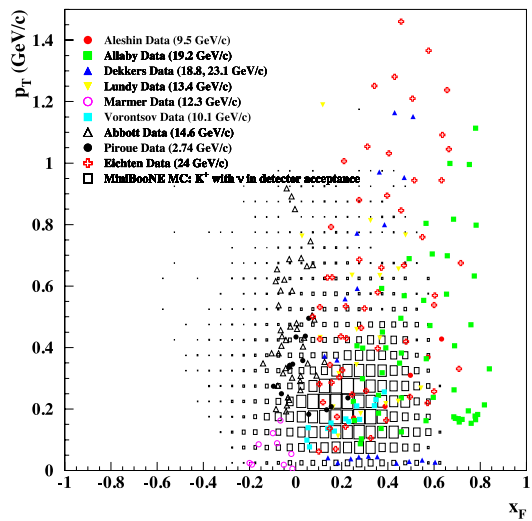}
\caption{Values of $x_{F}$ and $p_{T}$ for the data points of the various data
sets in Table~\ref{Data Sets}.  The distribution for kaons that produce
$\nu_{e}$ events in the MiniBooNE detector is shown as open boxes.}%
\label{xf_vs_pt_plot}%
\end{figure}

Several $K^{+}$ production measurements have been made for beam momentum less
than 25~GeV/c and are reported in Table~\ref{Data Sets}.
Those experiments, except for Piroue, have beam momenta higher than the BNB value of 8.89 GeV/c
although some of them such as Aleshin and Vorontsov are fairly close to the BNB beam momentum.
The kaons that produce neutrinos in MiniBooNE span the kinematic region with -0.1$<x_{F}<$0.5 and
0.05$<p_{T}(GeV/c)<$0.5 as shown in Figure~\ref{xf_vs_pt_plot}, which is nicely covered by the experimental
data sets listed in Table~\ref{Data Sets}. Of course, we are using the assumption that one can extrapolate these higher
beam momentum data to the BNB energy value using a parameterization such as
F-S. Thus, the first question to be answered is whether the data
appears to follow these scaling parameterizations.%

\begin{figure*}[hptb!]
\includegraphics[width=0.8\textwidth]{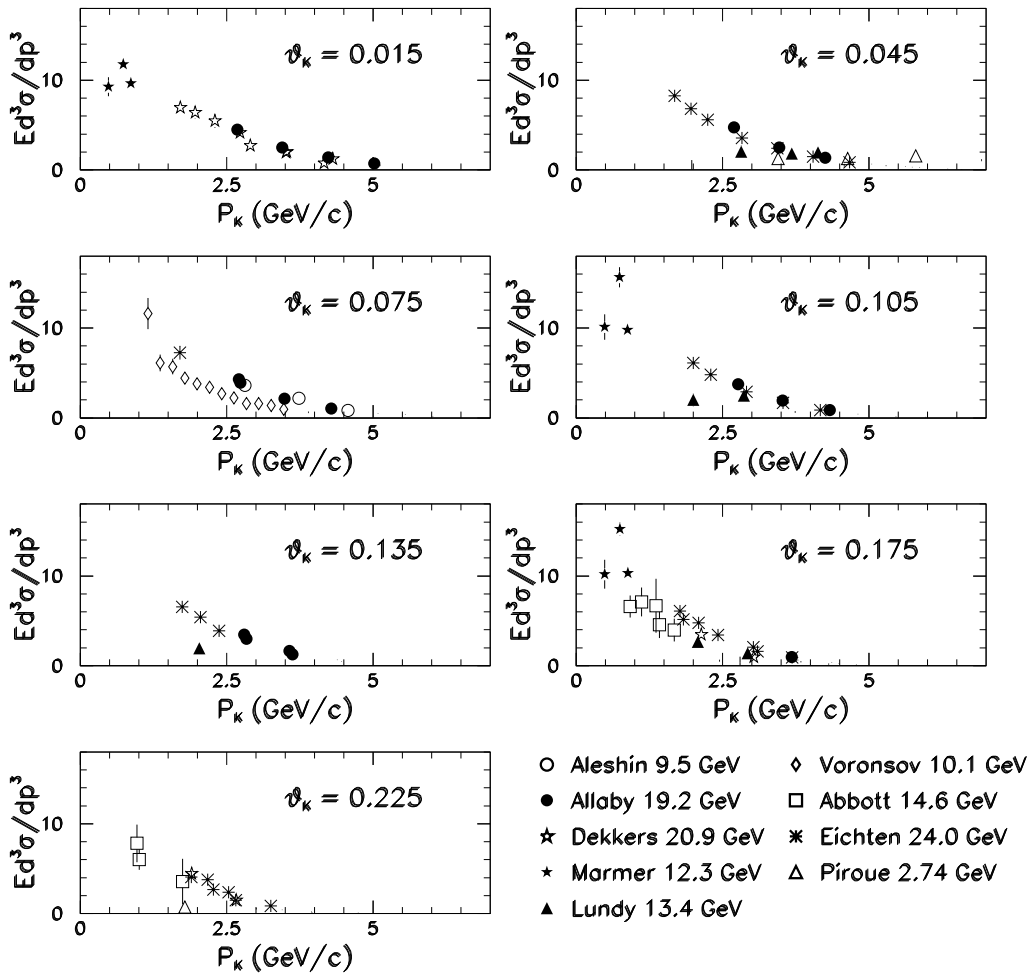}
\caption{$K^{+}$ production data sets scaled to the MiniBooNE beam momentum of 8.89~GeV/c using F-S.
The Y-axis units are ($mb \times c^3/GeV^2$). The production angle varies from 0 to 0.225 $rad$.}%
\label{fig:datasets}%
\end{figure*}

The F-S hypothesis says that the invariant cross section
$E\frac{d^{3}\sigma}{dp^{3}}$ should only depend on $x_{F}$ and $p_{T}$. This
hypothesis can further be tested by scaling all the data to a common beam momentum
and checked by the behavior of the invariant cross section against the scaled value of $p_{K}$ and
$\theta_{K}$. Figure~\ref{fig:datasets} shows the invariant cross section for scaled kaon momentum and angle bins using
the F-S assumption. For this plot, the data from each data set
is converted first to $x_{F}$ and $p_{T}$ and then scaled to $p_{K}^{8.89}$ and $\theta_{K}^{8.89}$
for a 8.89~GeV/c beam momentum.
For example, given a cross section point at $P_{BEAM}$=20~GeV/c with a given $P_K$ and $\theta_K$, one can calculate the $x_F$
and $p_T$ for this point. One can then find the equivalent $p'_K$ and $\theta'_K$ that would have the same $x_f$ and $p_T$ at $P_{BEAM}$=8.89~GeV/c.
As seen from the plots, the data appears to obey the scaling
hypothesis reasonably well except for the Lundy, Piroue, and Vorontsov data sets. Due to
the disagreements of the Lundy and Piroue data, these data sets are not
included in the fits described below. The Vorontsov data appears to
agree in shape with the other data sets but has an anomalous
normalization. Data sets not included in the fits are not
discarded. They are compared
separately to the fit results, as explained below.

\section{Feynman Scaling and Sanford-Wang Model Fits to the K$^{+}$ External Data Sets}\label{sec:fit_machinery}

Under the assumption that the experimental data follow the Feynman or S-W scaling models,
we can determine a parameterization that best fits these data sets. The
various production data sets are used as input to a fit for the scaling
function parameters that best describe the data. The fit uses a $\chi^{2}$
minimization technique using Minuit~\cite{minuit} to perform the numerical
minimization. Each experiment is allowed to have an independent normalization
parameter that is constrained by the published normalization uncertainty. The
fit minimizes the following function for an experiment $j$:

\begin{eqnarray}
\chi_{j}^{2}\ =\ \biggl[\sum_{i}\frac{(N_{j}\times SF_{i}-Data_{i})^{2}%
}{\left(  f\times\sigma_{i}\right)  ^{2}}\biggr]+\frac{(1-N_{j})^{2}}%
{\sigma_{N_{j}}^{2}},%
\end{eqnarray}

\noindent where $i$ is the ($P_{K}$,$\theta_{K}$) bin index, $SF$ is the scaling
function prediction evaluated at the given $(p_{BEAM},\theta_{K},p_{K})$,
$Data_{i}$ is the measurement at a given $(p_{BEAM},\theta_{K},p_{K})$,
$\sigma_{i}$ is the data error for measurement $i$, $f$ is the
scaling factor to bring the $\chi^{2}/d.o.f.=1,$ $N_{j}$ is the normalization
factor for experiment $j$, $\sigma_{N_{j}}$ is the normalization
uncertainty for experiment $j$, and d.o.f. indicates degree of freedom. The total $\chi^{2}$ for external data sets is
then the sum over the experiments of the individual $\chi^{2}_j$ values,

\begin{eqnarray}
\chi^{2}=\sum_{j}\chi_{j}^{2}.
\end{eqnarray}

The $\chi^{2}$ is minimized in order to obtain the best
values and uncertainties for the parameterization coefficients $c_{j}$, given in
Eq.~\ref{Scaling_Function} (or \ref{SW_Function},
and for the normalization factors $N_{j}$).
The uncertainties on the fit values at 1$\sigma$ are determined from a $\Delta \chi^2= 1$ change with respect to $\chi^2_{min}$ and the fit also yields a covariance matrix that can be used to propagate correlated errors associated with the parameterization of the cross section.

A F-S fit to all the experimental data sets with 0.0$<P_{K}^{8.89}(GeV/c)<$6.0 gives a $\chi^{2}
$/d.o.f. equal to 4.03 with large $\chi^{2}$ contributions from data
with $P_{K}^{8.89}<$1.2~GeV/c and $P_{K}^{8.89}>$5.5~GeV/c.
Therefore for the final scaling fits, the points with the larger pull terms, defined as ($(N_{j}\times SF_{i}-Data_{i})/\sigma_{i}$),
have been eliminated by only using data with 1.2$<P_{K}^{8.89}(GeV/c)<$5.5.

The 1.2~GeV/c cut effectively removes data at
negative $x_{F}$ where the nuclear environment starts to play an important
role. This cut also eliminates all the Marmer data points.

\begin{figure}[htbp!]
\begin{center}
\includegraphics[width = 0.8\columnwidth]{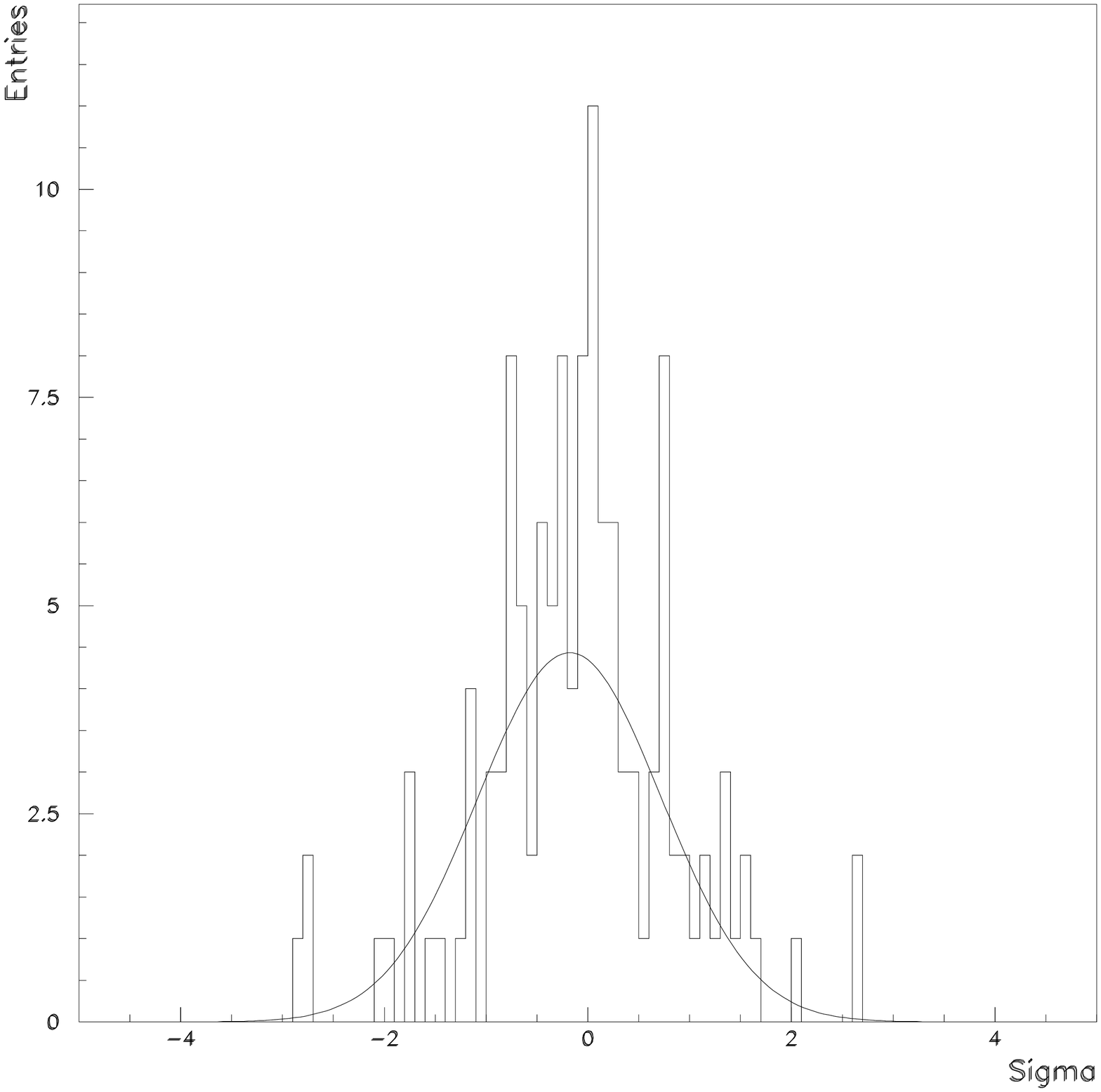}%
\caption{Values of the pull terms, $(N_{j}\times SF_{i}-Data_{i})/\sigma_{i}$, for
each data point for the F-S fit for 1.2$<P^{8.89}_{K}(GeV/c)<$5.5. The data
errors, $\sigma_{i}$, have been scaled up by $\sqrt{\chi^{2}/d.o.f.}=\sqrt{2.28} = 1.51$. The Gaussian fit gives a $\chi^2/number~of ~degree~of~freedom$ = 35.51/35, with a mean value = ($-0.18\pm0.11$) and sigma = ($0.90\pm0.17$).}%
\label{data_pulls_7par}%
\end{center}
\end{figure}

With all of these requirements, the $\chi^{2}$/d.o.f. for the F-S fit is reduced to 2.28. The uncertainties for the fitted
cross section need to be corrected for this $\chi^{2}$/d.o.f., which is larger than
1.0. This is accomplished by scaling up the errors of each of the data points by
$\sqrt{\chi^{2}/d.o.f.}$ before doing the fit.
Figure~\ref{data_pulls_7par} shows the pull terms for the
seven parameter F-S fit where the errors have been
scaled up by this $\sqrt{\chi^{2}/d.o.f.}=\sqrt{2.28}=$1.51.

\par A S-W fit to all the experimental data has been performed as well. To be able to directly compare the S-W with the F-S fit we
have included in the S-W fit only data with 1.2$<P_{K}^{8.89}(GeV/c)<$5.5. The $\chi^{2}$/d.o.f. for the S-W fit is equal to 6.05.

\begin{table}[htbp!]
\begin{center}
\begin{ruledtabular}
\begin{tabular}[c]{r|cc|c}
Feynman Scaling & \multicolumn{2}{l|}{1.2$<P^{8.89}_{K}(GeV/c)<$5.5} & \\
Fit & Value & Error &  \\\hline
c1 & 11.70 & 1.05 & \\
c2 & 0.88 & 0.13 & \\
c3 & 4.77 & 0.09 & \\
c4 & 1.51 & 0.06 & \\
c5 & 2.21 & 0.12 & \\
c6 & 2.17 & 0.43 & \\
c7 & 1.51 & 0.40 & Input Error\\\hline
Aleshin & 1.09 & 0.07 & 0.10\\
Allaby & 1.04 & 0.07 & 0.15\\
Dekkers & 0.84 & 0.06 & 0.20\\
Vorontsov & 0.53 & 0.04 & 5.00\\
Abbott & 0.76 & 0.07 & 0.15\\
Eichten & 1.00 & 0.07 & 0.15\\\hline
$\chi^{2}$/d.o.f. (no $f$) & 2.28 & ($d.o.f. = 119$)  & \\
\end{tabular}
\caption{Results for the F-S fits to the $K^+$ data including a
single normalization factor for each experiment. The data errors
have been scaled up by a factor of $\sqrt{\chi^2/d.o.f.} = $f$ = 1.51$ when included in the fit but the $\chi^2$/d.o.f. value listed is for the data without this scaling. d.o.f. indicates here degree of freedom and "no $f$" means no correction factor applied.}
\label{Fit_Results_FS}%
\end{ruledtabular}
\end{center}
\end{table}

\begin{table}[htbp!]
\begin{center}
\begin{ruledtabular}
\begin{tabular}[c]{r|cc|c}
Modified S-W & \multicolumn{2}{l|}{1.2$<P^{8.89}_{K}(GeV/c)<$5.5} & \\
Fit & Value & Error &  \\\hline
c1 & 14.89   & 1.89 & \\
c2 & 0.91     & 0.13 & \\
c3 & 12.80   & 7.46 & \\
c4 & 2.08     & 0.35 & \\
c5 & 2.65     & 0.50 & \\
c6 & 4.61     & 0.10 & \\
c7 & 0.26     & 0.01 & \\
c8 & 10.63   & 7.06 & \\
c9 & 2.04     & 0.01 & Input Error \\\hline
Aleshin        & 1.02 & 0.09 & 0.10 \\
Allaby          & 0.74 & 0.09 & 0.15 \\
Dekkers      & 0.57 & 0.08 & 0.20 \\
Vorontsov   & 0.42 & 0.04 & 5.00 \\
Abbott         & 1.38 & 0.11 & 0.15 \\
Eichten       & 0.59 & 0.08 & 0.15 \\\hline
$\chi^{2}$/df (no $f$) & 6.05 & ($d.o.f. = 117$) & \\
\end{tabular}
\end{ruledtabular}
\caption{Results for the S-W scaling fits to the $K^+$ data including a
single normalization factor for each experiment. The data errors
have been scaled up by a factor of $\sqrt{\chi^2/d.o.f.} = $f$ = 2.46$ when included in the fit but the $\chi^2$/d.o.f. value listed is for the data without this scaling. d.o.f. indicates here degree of freedom and "no $f$" means no correction factor applied.}
\label{Fit_Results_SW}%
\end{center}
\end{table}

\section{Comparison of Feynman Scaling to Sanford Wang Results and Neutrino Predictions}\label{sec:comparison}

\begin{figure*}[hptb!]
\includegraphics[width=0.8\textwidth]{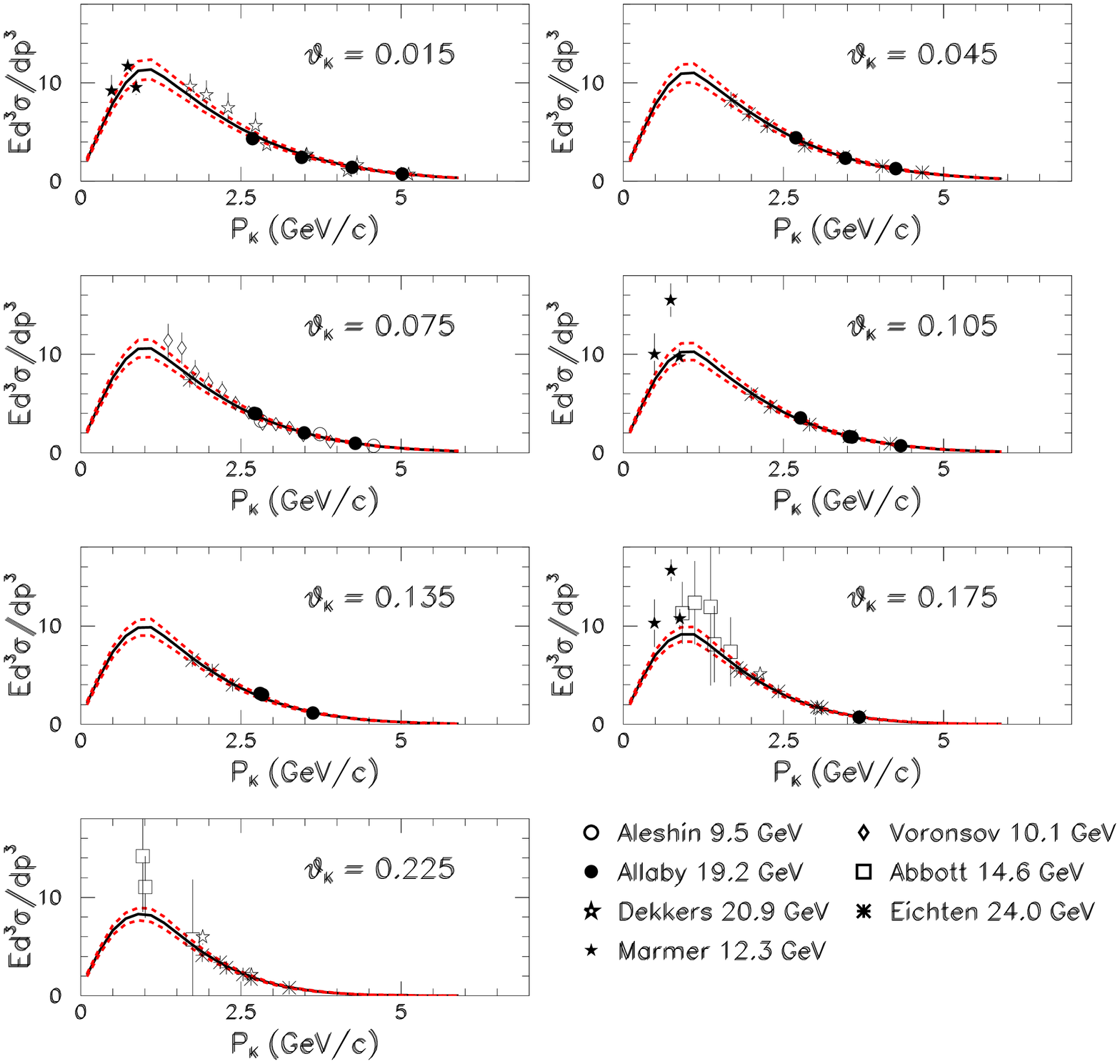}
\caption{Invariant kaon production cross section in $mb \times c^3/GeV^2$ versus kaon momentum for all
data along with the results of the F-S fit to data with 1.2$<P_{K}^{8.89}GeV/c<$5.5. The
$P_{K},\theta_{K},$ and invariant cross section fits and the data points
have been scaled to a beam momentum of 8.89~GeV/c assuming F-S and
normalized according to the fit results. This plot shows data and fit results for various value of $\theta$ in bins
from 0 to 0.225~rad. The three solid curves show the central value and $1\sigma$ uncertainty for the
F-S fit.}%
\label{fig:fit_result_FS}%
\end{figure*}

\begin{figure*}[hptb!]
\includegraphics[width=0.8\textwidth]{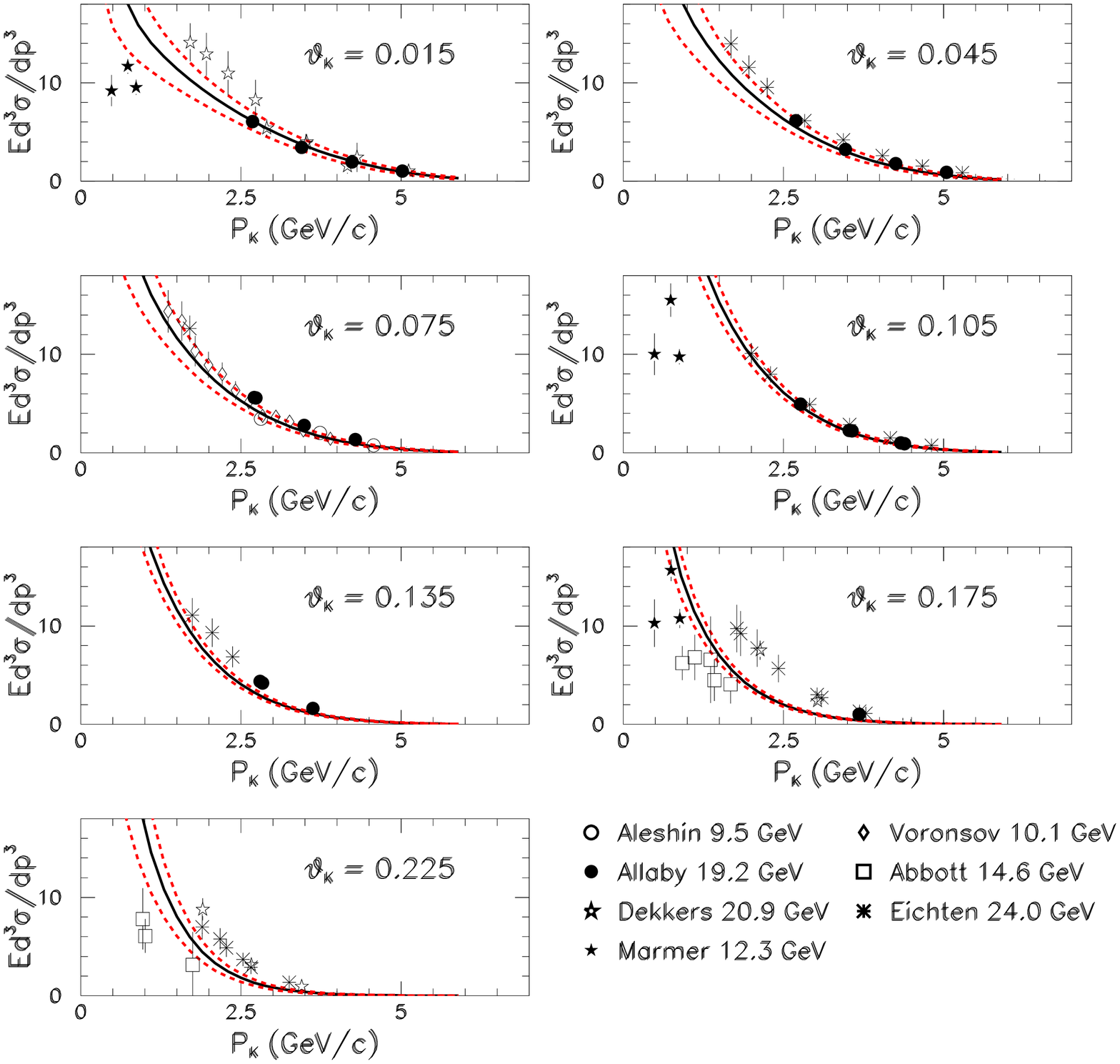}
\caption{Invariant kaon production cross section in $mb \times c^3/GeV^2$ versus kaon momentum for all
data along with the results of the S-W scaling fit to data with 1.2$<P_{K}^{8.89}GeV/c<$5.5. The
$P_{K},\theta_{K},$ and invariant cross section fits and the data points
have been scaled to a beam momentum of 8.89~GeV/c assuming F-S and
normalized according to the fit results. This plot shows data and fit results for various value of $\theta$ in bins
from 0 to 0.225~rad. The three solid curves show the central value and $1\sigma$ uncertainty for the
S-W scaling fits.}%
\label{fig:fit_result_SW}%
\end{figure*}

Tables~\ref{Fit_Results_FS} and \ref{Fit_Results_SW} report the final fit values for the coefficients
and the normalization factors for the F-S and S-W parameterizations respectively.
Figures~\ref{fig:fit_result_FS} and \ref{fig:fit_result_SW}
show the fit function curves for the F-S and S-W parameterizations as compared to the data. The fits are stable with respect to parameter starting values and yield positive definite covariance matrices.
The error bands in Figures~\ref{fig:fit_result_FS} and \ref{fig:fit_result_SW} are determined by
propagating the covariance matrix for the $c_{j}$ parameters to
the invariant cross section errors.

As seen from the plots, the F-S function gives a very good description of the data over the full kaon momentum range used in the fit and has a  reasonable $\chi^{2}$/d.o.f. = 2.3.
Below 1.2~GeV/c, the F-S prediction has some disagreement with a few the of the Marmer (not included in the fit) and Abbott data points but in general is also fitting that well in that region. The normalization factors for the F-S fits are within $1\sigma$ of the quoted experimental error except for the Vorontsov data (see Table~\ref{Fit_Results_FS}).  As mentioned above, the Vorontsov data shows a systematically low normalization with respect the other sets of scaled data.
Therefore, for all the scaling fits, the Vorontsov data has only been used for shape information
by giving the normalization a large uncertainty (500\%).

In contrast, the S-W final fit parameterization has rather large discrepancies with the data in almost all regions and has a much larger $\chi^{2}$/d.o.f. = 6.05.  Additionally, the normalization factors given in Table~\ref{Fit_Results_SW} are very much outside of the quoted experimental errors and, for example, the factors for Eichten and Allaby differ from 1.0 by 2 to 3$\sigma$.

Tables~\ref{Sweet_Spot_xsecs_FS} and \ref{Sweet_Spot_xsecs_SW} list the differential cross sections for several
different kinematic points for kaon production. The uncertainties are
obtained by propagating the covariance matrix for the $c_{j}$ coefficients
into the scaling function. The first three points in Table~\ref{Sweet_Spot_xsecs_FS} and \ref{Sweet_Spot_xsecs_SW}
correspond to the mean kaon production points that produce electron neutrino of 0.35, 0.65, and
0.95~GeV in MiniBooNE. The fourth point corresponds to the kaon kinematics that produce average energy neutrinos from all kaon decays (called the
\textquotedblleft kaon sweet spot\textquotedblright), and the fifth point is
associated with the mean kaon kinematics for the highest energy kaon-decay muon neutrinos observed in MiniBooNE. As seen from Table~\ref{Sweet_Spot_xsecs_FS} and \ref{Sweet_Spot_xsecs_SW} ,
the two parameterizations give much different results for the cross section values and uncertainties with the F-S fit giving a larger value by a factor 2 for the lowest energy neutrino bin at 0.35 GeV. The source of the this discrepancy is a large drop in the invariant cross section of the S-W parameterization at large angles.

\begin{table}[htbp!]
\begin{center}
\begin{ruledtabular}
\begin{tabular}
[c]{c|cc|c}
 & $P^{8.89}_{K}$ & $\theta_{K}$ & $\sigma_{K\text{ }prod}$ \\
 & (GeV/c)  & (rad)  & (mb)  \\\hline
$E_{\nu}=0.35$ GeV & 1.52 & 0.213 &   ~9.37 $\pm$ 0.73 (7.8\%)\\
$E_{\nu}=0.65$ GeV & 2.07 & 0.127 & 10.69 $\pm$ 0.75 (7.0\%) \\
$E_{\nu}=0.90$ GeV & 2.45 & 0.103 & 10.22 $\pm$ 0.71 (6.9\%) \\
Kaon Sweet Spot & 2.80 & 0.106 & ~8.67 $\pm$ 0.60 (6.9\%) \\
HE $\nu_\mu$ Events & 4.30 & 0.055 & ~4.73 $\pm$ 0.33 (7.0\%) \\
\end{tabular}
\caption{Differential cross section values for various kinematic points for the 1.2$<P_{K}<$5.5~GeV/c F-S fit.
The first three results are for the average kaon kinematics that give electron neutrinos with the
given energy. The last
result is for the average kaon kinematics associated with highest energy $\nu_{\mu}$ events in MiniBooNE.}
\label{Sweet_Spot_xsecs_FS}%
\end{ruledtabular}
\end{center}
\end{table}

\begin{table}[htbp!]
\begin{center}
\begin{ruledtabular}
\begin{tabular}
[c]{c|cc|c}
 & $P^{8.89}_{K}$ & $\theta_{K}$ & $\sigma_{K\text{ }prod}$ \\
 & (GeV/c)  & (rad) & (mb)  \\\hline
$E_{\nu}=0.35$ GeV & 1.52 & 0.213 & 4.25 $\pm$ 0.77 (18\%) \\
$E_{\nu}=0.65$ GeV & 2.07 & 0.127 & 8.99 $\pm$ 1.34 (15\%) \\
$E_{\nu}=0.90$ GeV & 2.45 & 0.103 & 9.91 $\pm$ 1.43 (14\%) \\
Kaon Sweet Spot & 2.80 & 0.106 	 & 7.73 $\pm$ 1.13 (15\%) \\
HE $\nu_{\mu}$ Events & 4.30 & 0.055 & 5.24 $\pm$ 0.84 (16\%) \\
\end{tabular}
\caption{Differential cross section values for various kinematic points for the 1.2$<P_{K}<$5.5~GeV/c S-W scaling fit.
The first three results are for the average kaon kinematics that give electron neutrinos with the
given energy. The fourth result is the previous point used for a kaon sweet spot. The last
result is for the average kaon kinematics associated with highest energy $\nu_{\mu}$ events in MiniBooNE.}
\label{Sweet_Spot_xsecs_SW}%
\end{ruledtabular}
\end{center}
\end{table}

\begin{figure*}[htbp!]
\begin{center}
\includegraphics[width = 0.8\textwidth]{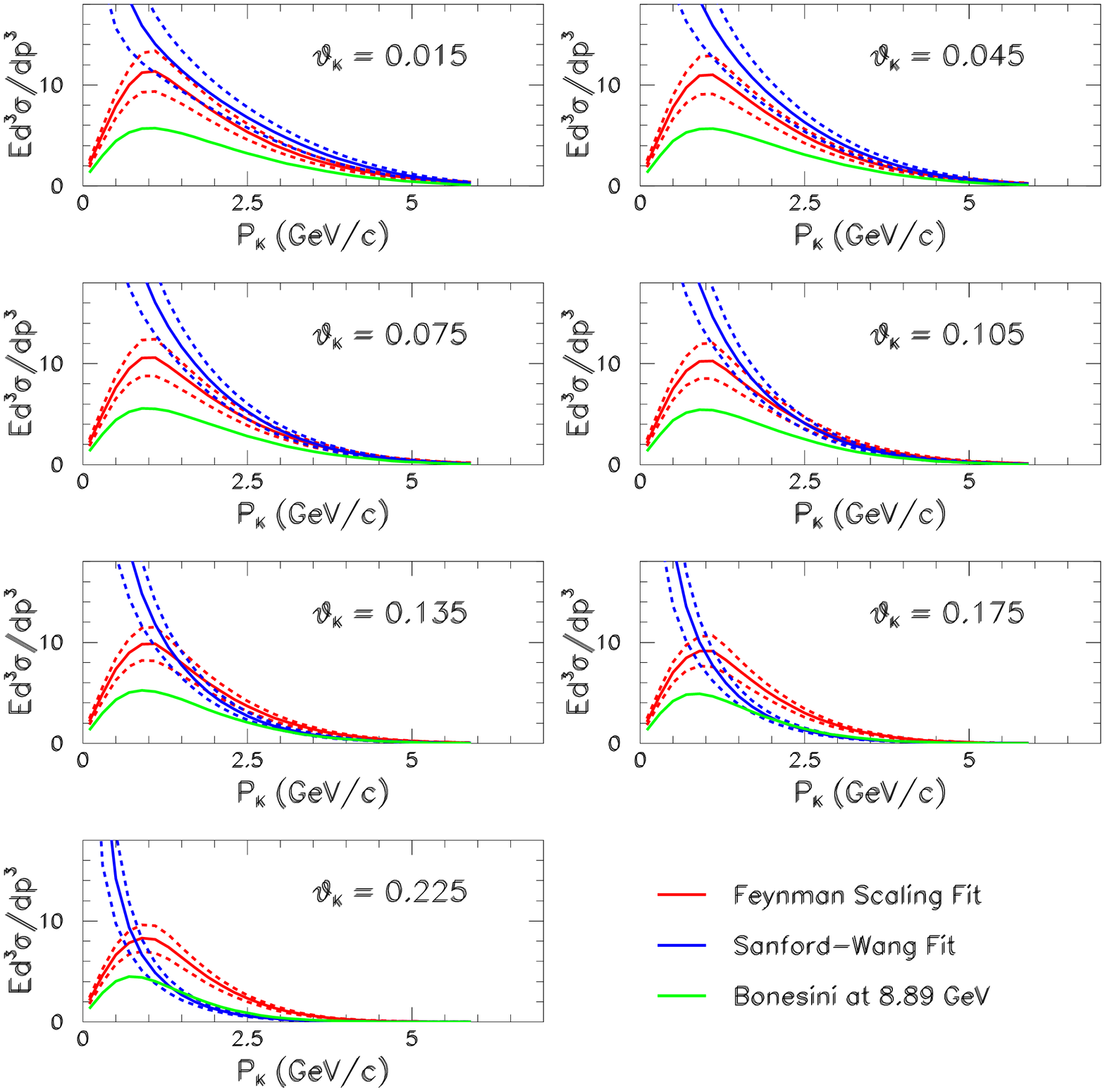}

\caption{Invariant kaon production cross section in units of $mb \times c^3/GeV^2$ versus kaon momentum in GeV/c for the S-W, F-S and radial scaling (Bonesini)\cite{Bonesini:2001iz} parameterizations for a beam momentum of 8.89~GeV/c. The results are shown for various $\theta$ bins from 0 to 0.225~rad. The three solid curves, respectively for the F-S and S-W fits, show the central value and 1$\sigma$ uncertainty for each of the fits.}

\label{fig:FS_SW_BO}
\end{center}
\end{figure*}

The predictions for the size and kinematic dependence of the invariant differential cross section as function of $K^+$ momentum are quite different for the F-S and S-W parameterizations as shown in Figure~\ref{fig:FS_SW_BO}, especially for low value of the $K^+$ momentum.

To illustrate the difference between the F-S and the S-W
predictions, we have used an analytic simulation of the BNB neutrino beam line
designed for the MiniBooNE experiment (described in Reference~\cite{AguilarArevalo:2008yp}).
Table~\ref{Comparison_Table} gives the comparison of the
predicted $\nu_{e}$ event rate from $K^{+} \rightarrow \pi e^+ \nu_{e}$ using the
above F-S and S-W production parameterizations as calculated using
this BNB simulation.

\begin{table*}[htbp!]
\begin{center}
\begin{ruledtabular}
\begin{tabular}
[c]{c|ccrrrr}
$\theta_K$ & \multicolumn{3}{c}{$K_{e3}^{+}$ {Feynman Scaling Fit}} & \multicolumn{3}{|c}{$K_{e3}^{+}$ {Sanford-Wang Fit}}\\
{Angular Bins(rad) } & {All }E$_{\nu}$(GeV) & $<$1~GeV &
\multicolumn{1}{c}{$>$2~GeV} & \multicolumn{1}{|c}{{All }$E_{\nu}$(GeV)} & \multicolumn{1}{c}{$<$1~GeV} & \multicolumn{1}{c}{$>$2~GeV}\\\hline
{  0.015} & {  36.7} & {  2.6} & \multicolumn{1}{c}{{  18.0}} & \multicolumn{1}{|c}{{  43.4}} & \multicolumn{1}{c}{{  3.3}} & \multicolumn{1}{c}{{  19.3}}\\
{  0.045} & {  92.5} & {  8.4} & \multicolumn{1}{c}{{  35.9}} & \multicolumn{1}{|c}{{  111.0}} & \multicolumn{1}{c}{{  12.0}} & \multicolumn{1}{c}{{  35.9}}\\
{  0.075} & {  110.5} & {  13.7} & \multicolumn{1}{c}{{  27.0}} & \multicolumn{1}{|c}{{  141.3}} & \multicolumn{1}{c}{{  22.6}} & \multicolumn{1}{c}{{  26.5}}\\
{  0.105} & {  96.8} & {  17.2} & \multicolumn{1}{c}{{  4.4}} & \multicolumn{1}{|c}{{  138.3}} & \multicolumn{1}{c}{{  32.6}} & \multicolumn{1}{c}{{  4.1}}\\
{  0.135} & {  59.1} & {  21.8} & \multicolumn{1}{c}{{  0.0}} & \multicolumn{1}{|c}{{  100.5}} & \multicolumn{1}{c}{{  45.8}} & \multicolumn{1}{c}{{  0.0}}\\
{  0.175} & {  39.4} & {  32.4} & \multicolumn{1}{c}{{  0.0}} & \multicolumn{1}{|c}{{  83.7}} & \multicolumn{1}{c}{{  73.9}} & \multicolumn{1}{c}{{  0.0}}\\
{  0.225} & {  21.9} & {  21.9} & \multicolumn{1}{c}{{  0.0}} & \multicolumn{1}{|c}{{  56.8}} & \multicolumn{1}{c}{{  56.8}} & \multicolumn{1}{c}{{  0.0}}\\
{  Total} & {  476.6} & {  137.9} & \multicolumn{1}{c}{{  85.3}} & \multicolumn{1}{|c}{{  731.2}} & \multicolumn{1}{c}{{  303.4}} & \multicolumn{1}{c}{{  85.9}}\\
\end{tabular}
\caption{Electron neutrino event rate in MiniBooNE for 5.0$\times 10^{20}$ proton on target for $K^{+}_{e3}$ decays with
F-S and S-W parameterizations. The events were calculated using MiniBooNE simulation and are for a beam radius
less than 6.0~m. The different columns list the selected electron neutrino events for all $E_\nu$, $E_\nu<$ 1~GeV, and $E_\nu>$ 2~GeV. Uncertainty in the neutrino event rate due to the F-S or S-W parametrization is 7\% and 15\% respectively as described in Table~\ref{Sweet_Spot_xsecs_FS} and \ref{Sweet_Spot_xsecs_SW}.}
\label{Comparison_Table}
\end{ruledtabular}
\end{center}
\end{table*}

\section{High energy parameterization}

The hypothesis of F-S has also been verified to hold with different parameterizations over a wide range of primary proton beam energies (from 24~GeV to 450~GeV). In Bonesini {\it et al.} \cite{Bonesini:2001iz} data at higher proton energies has been empirically parameterized as a function of the transverse momentum ($p_T$) and the scaling variable $x_R = E^*/E^*_{max}$ where $E^*$ is the energy of the particle in center-of-mass frame. The choice of these variables for the description of the invariant cross section (radial scaling) is motivated again by an assumed scaling behavior of the invariant cross section. The radial scaling variable is approximately equal to the F-S variable at high energy and has the property of never taking on a negative value. (A detailed comparison of radial scaling and F-S can be found in \cite{taylor:1976,Norbury:2009}, where the authors compare different models with the production data at different energies down to about 24 GeV.)

Bonesini {\it et al.}\cite{Bonesini:2001iz} has obtained an empirical parameterization based on radial scaling fits to data collected with 400~GeV/c and 450~GeV/c protons incident on a Be target. The results from this parameterization are compared in Figure~\ref{fig:FS_SW_BO} to the predictions of F-S and S-W models at a proton momentum of 8.89 GeV/c. As seen from Figure~\ref{fig:FS_SW_BO}, this radial scaling model underestimates $K^+$ production at a beam momentum of 8.89 GeV/c by more than a factor of two even though the parameterization describes well the high proton momentum data  ($>$24~GeV/c)~\cite{Bonesini:2001iz}.


\section{The SciBooNE $K^+$ Measurements}\label{sec:sciboone_meas}

The SciBooNE collaboration has reported a measurement~\cite{Cheng:2011wq} for $K^+$ production in the BNB with respect to the Monte Carlo (MC) beam simulation.
The SciBooNE experiment collected data in 2007 and 2008 with neutrino ($0.99\times 10^{20}$ protons on target (POT)) and
antineutrino ($1.53\times10^{20}$ POT) beams in the FNAL BNB line.
The SciBooNE detector is located 100~m downstream from the neutrino production target.
The flux-averaged mean neutrino energy is 0.7~GeV in neutrino running mode and 0.6~GeV in antineutrino running mode
\par The SciBooNE detector consists of three detector components; SciBar, Electromagnetic
Calorimeter (EC) and Muon Range Detector (MRD). SciBar is a fully active and
fine grained scintillator detector that consists of 14,336 bars arranged in vertical and
horizontal planes. SciBar is capable of detecting all charged particles and performing
dE/dx-based particle identification. The EC is located downstream of SciBar. The detector
is a ÒspaghettiÓ calorimeter with thickness of 11 radiation lengths and is used to measure $\pi^0$
and the intrinsic $\nu_e$ component of the neutrino beam. The MRD is located downstream
of the EC in order to measure the momentum of muons up to 1.2~GeV/c with range.
It consists of 2-inch thick iron plates sandwiched between layers of plastic scintillator
planes.

\par In the SciBooNE experiment, particle production is simulated using the methods described in Ref.~\cite{AguilarArevalo:2008yp}. The production of $K^{+}$ is simulated using the F-S formalism as described in Section~\ref{sec:FSformalism} with the coefficients reported in Table~\ref{Fit_Results_FS}. The predicted double differential cross section at the mean momentum and angle for kaons which produce neutrinos in SciBooNE ($p_K$ = 3.87~GeV/c and $\theta_K$ = 0.06~rad) is

\begin{eqnarray}
\dfrac{d^2\sigma}{dpd\Omega}~= ~(6.3\pm0.44)~mb/(GeV/c \times sr),
\label{eq:central_value_doublediff}
\end{eqnarray}

The error on the double differential cross section prediction using the F-S parametrization at the SciBooNE $p_K$ and $\theta_K$ is 7\%. The SciBooNE and MiniBooNE collaboration have adopted a conservative error of 40\%. This larger error was chosen because of the uncertainties in extrapolating the $K^+$ prediction data from high to low proton beam energy using the F-S and S-W models as explained in References~\cite{Cheng:2011wq,AguilarArevalo:2008yp}.

\subsection{SciBooNE $K^+$ Production Measurement}\label{sec:sciboone_rate}

The SciBooNE data can be used as an additional constraint in fits to $K^+$ production cross sections. In SciBooNE, neutrinos from $K^+$ decay are selected using high energy $\nu_\mu$ interactions within the volume of the SciBar detector. The high-energy selection is accomplished by isolating charged current interactions
that produce a muon that crosses the entire MRD. This sample is further divided into three sub-samples based on whether 1, 2, or 3 reconstructed SciBar tracks are identified at the neutrino interaction vertex in the SciBar detector. Since the reconstruction of the energy of the muon is not possible because the muon exits the MRD detector, the reconstructed muon angle relative to beam axis is used as the primary kinematic variable to separate neutrinos from pion and kaon decay.
The values for $\dfrac{d^2\sigma}{dpd\Omega}$ for neutrino, antineutrino, and combined data mode running are given in Table~\ref{tab:neutrino+antineutrino_energy_angle_NUANCE} along with the mean energy and angles for the corresponding $K^+$ samples. The F-S and S-W prediction values are obtained using the parametrizations described in Section~\ref{sec:FSformalism} and~\ref{sec:SWformalism} along with the parameters listed in Table~\ref{Fit_Results_FS} and~\ref{Fit_Results_SW}. 

\begin{table}[htbn!]
\begin{center}
\vspace{0.5cm}
\begin{ruledtabular}
\begin{tabular}{c|c|c|c}
                                                              & $E_{K^+}$ (GeV)   & $\theta_{K^+}$(rad)       & $\dfrac{d^2\sigma}{dpd\Omega}$ \\
                                                              &                               &                                        & (mb/(GeV/c $\times$ sr) \\ \hline
$\nu$-mode                                           & 3.81$\pm$0.03          & 0.07$\pm$0.01                   &  5.77$\pm$0.83      \\
$\bar{\nu}$-mode                                  & 4.29$\pm$0.06          & 0.03$\pm$0.01                   &  3.18$\pm$1.94       \\
$\nu$ + $\bar{\nu}$-mode                     & 3.90$\pm$0.03          & 0.06$\pm$0.01                   &  5.34$\pm$0.76       \\\hline
F-S prediction					     & 3.90				 & 0.06				       &  6.30$\pm$0.44(7\%)       \\
S-W prediction				             & 3.90				 & 0.06				       &  6.84$\pm$1.09(16\%)       \\
\end{tabular}
\end{ruledtabular}
\caption{Measured $\dfrac{d^2\sigma}{dpd\Omega}$, mean energy, and mean angle (with respect to proton beam direction) for the selected $K^+$ in neutrino, antineutrino, and the combined neutrino and antineutrino samples using MiniBooNE MC. Errors on the mean energy and mean angle values correspond to the error on the mean for the relative distributions. F-S and S-W predictions are also reported at the mean SciBooNE $K^+$ energy and angle.}
\label{tab:neutrino+antineutrino_energy_angle_NUANCE}
\end{center}
\end{table}

\begin{figure}[htbp!]
\subfigure[Neutrino Sample]{\includegraphics[width = 0.8\columnwidth]{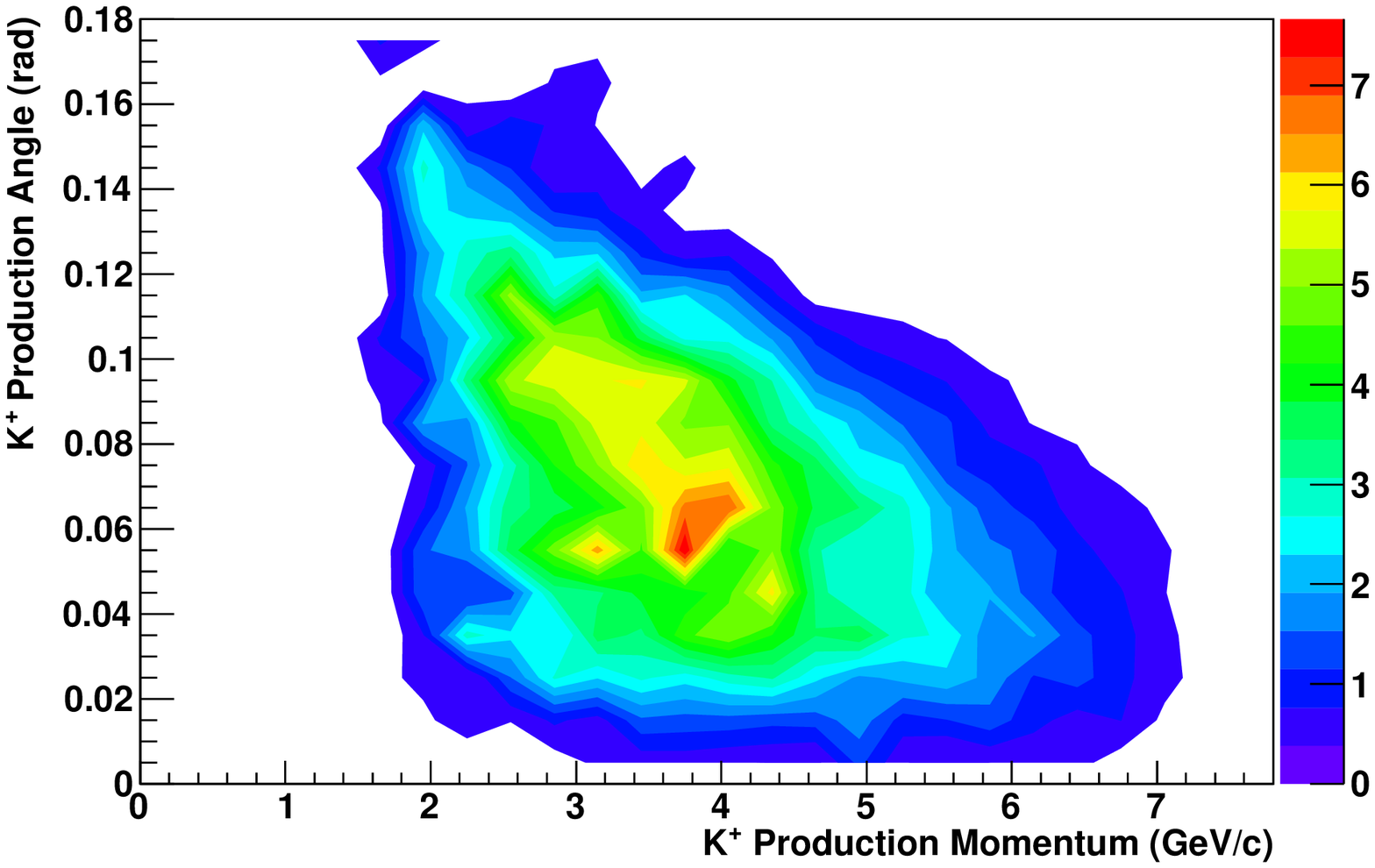}}
\subfigure[Antineutrino Sample]{\includegraphics[width = 0.8\columnwidth]{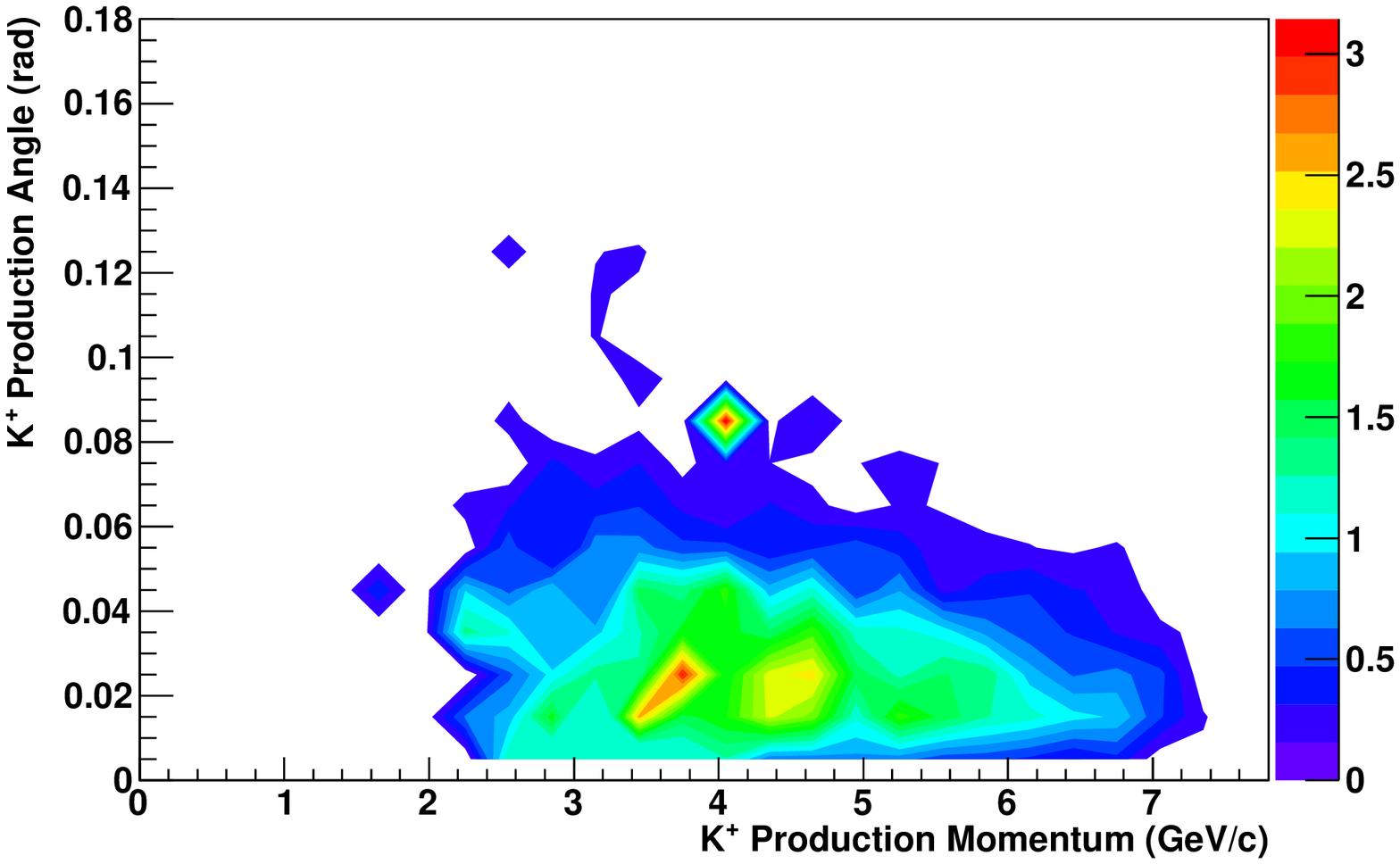}}
\caption{The true $K^+$ momentum versus angle distribution in the SciBooNE MC for neutrino-mode (on top) and antineutrino mode (on bottom) running. The unit for the color scale is number of events POT normalized.}
\label{fig:sciboone_nu_mode_E_vs_theta}
\end{figure}

\noindent The $K^{+}$ momentum versus angle distribution for the 2-track SciBar sample in the simulation is shown in Figure~\ref{fig:sciboone_nu_mode_E_vs_theta}.

\begin{figure}[htbp!]
\begin{center}
\includegraphics[width = 0.8\columnwidth]{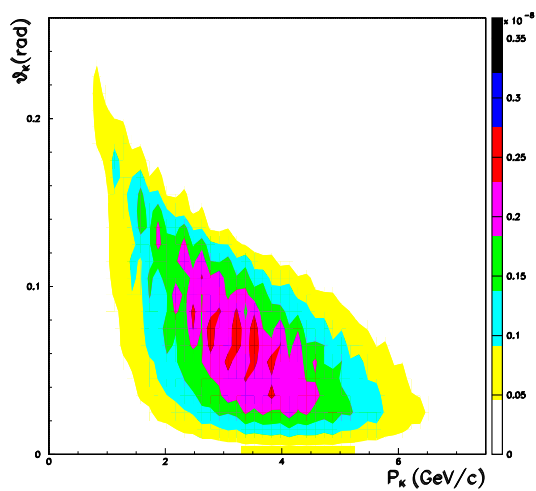}
\caption{Kinematical region as function of angle and momentum for the $K^{+}$ mesons that produce $\nu_{e}$ events in
MiniBooNE. The unit for the color scale is number of events.}
\label{fig:kaon_that_produce_nues}
\end{center}
\end{figure}

Figure~\ref{fig:sciboone_nu_mode_E_vs_theta} shows the kinematics of the selected $K^+$ events in SciBooNE, while Figure~\ref{fig:kaon_that_produce_nues} shows the kinematical region as function of angle and momentum for $K^+$ mesons that produce $\nu_e$ events in MiniBooNE.

The SciBooNE measurement is a direct test of the extrapolation of parameterizations found from higher beam energies to the MiniBooNE beam energy. The predictions for the double differential cross section for the F-S and S-W models are reported in Table~\ref{tab:neutrino+antineutrino_energy_angle_NUANCE} and shows a good agreement with the SciBooNE measurement, a better agreement is found in the case of the F-S parametrization.

The SciBooNE $K^+$ production measurements can also be added to the F-S fit as additional external data using the following procedure.
First we retrieve all the SciBooNE MC $K^+$ events with their $\theta_i$ and $p_i$ for the neutrino and antineutrino sample. Then we calculate the following quantities:

 \begin{eqnarray}
 N_i &=& \sum_i{ \frac{ \dfrac{d^2\sigma}{dpd\Omega}(c_{fit},\theta_i,p_i)} { \dfrac{d^2\sigma}{dpd\Omega}(c_{MC},\theta_i,p_i)}}\\
 N_0 &=& \sum_i{1}
 \end{eqnarray}

 \noindent These quantities are then used at each fit step to build a pull term, defined in Eq.~\ref{eq:SB_pull_term}, to be added to the $\chi^2$ of the fit.

\begin{eqnarray}
pull-term_{\nu,\bar{\nu}} = \frac{\left (\frac{N_i}{N_0}-K^+_{prod,SB}\right)}{\left(error~K^+_{prod,SB}\right)^2}
\label{eq:SB_pull_term}
\end{eqnarray}

Each data point in $\theta_i$ and $p_i$ is reweighted using the double differential cross section value for the current set of

$c_i$ coefficient of Eq.~\ref{Scaling_Function} computed at each step of the Minuit fit. The set of coefficient used in the MC is labeled as $c_{MC}$, the values of these coefficients is listed in Table~\ref{Fit_Results_FS}. The $K^+_{prod,SB}$ and $error~K^+_{prod,SB}$ in the Eq.~\ref{eq:SB_pull_term} are the values of the SciBooNE production measurement and error (see Table~\ref{tab:neutrino+antineutrino_energy_angle_NUANCE}), respectively.

Two separate pull terms are added to the fit $\chi^2$ corresponding to the SciBooNE neutrino and antineutrino $K^+$ production measurements.

\begin{figure}[hptb!]
\includegraphics[width=0.8\columnwidth]{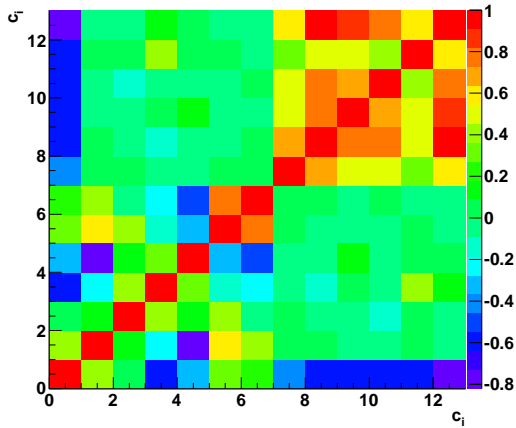}
\caption{Correlation for the seven parameters in the F-S fit function and six
normalization factor parameters after applying the SciBooNE constraint to the fit  due to the $K^+$ production measurement.}%
\label{fig:Correlation_Matrix_after_SB_prod}%
\end{figure}

The results of scaling function fit to all experiments with 1.2$<P^{8.89}_{K}<$5.5~GeV/c, including the SciBooNE data,
are given in Table~\ref{tab:Fit_Results_after_SB_prod}. The covariance matrix is given in Table~\ref{tab:Covariance_Matrix_after_SB_prod} and the
correlation matrix is presented in Figure~\ref{fig:Correlation_Matrix_after_SB_prod}.

Table~\ref{tab:Sweet_Spot_xsecs_after_SB_prod} lists the differential cross section for the kaon production at the various kaon kinematic points.
The uncertainties are obtained as described in Section~\ref{sec:comparison}.

\begin{table}[htbp!]
\begin{center}
\begin{ruledtabular}
\begin{tabular}[c]{r|cc|c}
Scaling Fits & \multicolumn{2}{l|}{1.2$<P^{8.89}_{K}(GeV/c)<$5.5} & \\
& Value & Error &  \\\hline
c1 & 11.29 & 0.93 & \\
c2 & 0.87   & 0.13 & \\
c3 & 4.75   & 0.09 & \\
c4 & 1.51   & 0.06 & \\
c5 & 2.21   & 0.12 & \\
c6 & 2.17   & 0.43 & \\
c7 & 1.51   & 0.40 & Input Error\\\hline
Aleshin & 1.12 & 0.07 & 0.10\\
Allaby & 1.07 & 0.06 & 0.15\\
Dekkers & 0.87 & 0.06 & 0.20\\
Vorontsov & 0.55 & 0.04 & 5.00\\
Abbott & 0.79 & 0.07 & 0.15\\
Eichten & 1.03 & 0.06 & 0.15\\\hline
$\chi^{2}$/df (no $f$) & 2.28 & ($d.o.f. = 119$) & \\
\end{tabular}
\end{ruledtabular}
\caption{Results for the F-S fits to the $K^+$ data including a
single normalization factor for each experiment and including the two SciBooNE pull term constraints.  Error treatment is the same as described in Section~\ref{sec:fit_machinery}. d.o.f. indicates here degree of freedom and "no $f$" means no correction factor applied.}
\label{tab:Fit_Results_after_SB_prod}%
\end{center}
\end{table}

\begin{table*}[htbp!]
\begin{center}
\begin{ruledtabular}
\begin{tabular}[c]{r|ccccccc}
& c1 & c2 & c3 & c4 & c5 & c6 & c7\\\hline
c1  & 0.84        &   0.48E-01 & 0.39E-02 & -0.32E-01 & -0.36E-01 & 0.12   &   0.69E-01 \\
c2  & 0.48E-01 & 0.16E-01  & 0.14E-02 & -0.15E-02 & -0.13E-01 & 0.32E-01 &  0.22E-01\\
c3  & 0.39E-02 & 0.14E-02  & 0.73E-02  & 0.20E-02  & 0.19E-02 & 0.14E-01 & -0.29E-02 \\
c4  & -0.32E-01 & -0.15E-02 & 0.20E-02  & 0.34E-02 & 0.20E-02 & -0.39E-02 & -0.60E-02 \\
c5  & -0.36E-01 & -0.13E-01 & 0.19E-02  & 0.20E-02 & 0.15E-01 & -0.15E-01 & -0.24E-01 \\
c6  & 0.12         & 0.32E-01  & 0.14E-01 & -0.39E-02 & -0.15E-01 & 0.18  &    0.12   \\
c7  & 0.69E-01 & 0.22E-01  & -0.29E-02 & -0.60E-02 & -0.24E-01 & 0.12 &      0.15 \\
\end{tabular}
\end{ruledtabular}
\caption{Covariance matrix for the seven scaling function fit parameters
after applying the SciBooNE production measurements in the F-S fit.}
\label{tab:Covariance_Matrix_after_SB_prod}
\end{center}
\end{table*}

\begin{table}[htbp!]
\begin{center}
\begin{ruledtabular}
\begin{tabular}
[c]{c|cc|c}
& $P^{8.89}_{K}$ & $\theta_{K}$ & $\sigma_{K\text{ }prod}$ \\
& (GeV/c)  & (rad) & (mb)  \\\hline
$E_{\nu}=0.35$ GeV & 1.52 & 0.213 &   ~9.05 $\pm$ 0.62 (6.9\%) \\
$E_{\nu}=0.65$ GeV & 2.07 & 0.127 & 10.32 $\pm$ 0.62 (6.0\%) \\
$E_{\nu}=0.90$ GeV & 2.45 & 0.103 &  ~9.87 $\pm$ 0.58 (5.9\%) \\
Kaon Sweet Spot & 2.80 & 0.106 & ~8.37 $\pm$ 0.49 (5.9\%) \\
HE $\nu_{\mu}$ Events & 4.30 & 0.055 & ~4.57 $\pm$ 0.27 (5.9\%) \\
\end{tabular}
\end{ruledtabular}
\caption{Differential cross section values for various kinematic points as in Table~\ref{Sweet_Spot_xsecs_FS} but including in the F-S fit the SciBooNE production measurement for neutrino
and antineutrino.}
\label{tab:Sweet_Spot_xsecs_after_SB_prod}%
\end{center}
\end{table}

Table~\ref{tab:Covariance_Matrix_after_SB_prod} gives the
covariance matrix for the baseline scaling fit using kaon
production data with 1.2$<P^{8.89}_{K}<$5.5~GeV/c. The correlation matrix is basically
made of two blocks, one associated with the $c_{1}$ through $c_{7}$ parameters
and one associated with the experimental normalization factors. The only
coupling of these two sets is through $c_{1}$ which has significant
correlations with the normalization factors. This is expected since the
$c_{1}$ parameter sets the normalization of the scaling function and should be
determined by the data normalizations.%

The terms of the covariance matrix from the F-S fit that includes the SciBooNE production measurement
include the factor 1.51 for the data set errors rescaling.

\begin{figure}[hptb!]
\includegraphics[width=0.8\columnwidth]{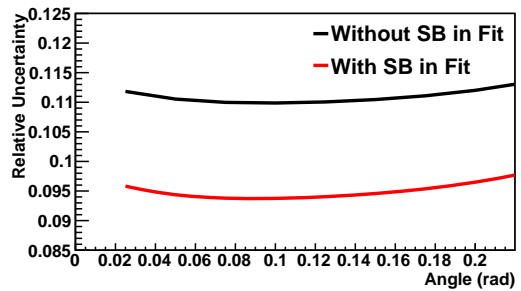}
\caption{Relative uncertainty on the double differential cross section as function of $K^+$ angle (0.0$<\theta_{K}<$0.25~rad) predicted by the F-S with and without including the SciBooNE production measurement.}%
\label{fig:unce_vs_angle}
\end{figure}

\begin{figure}[hptb!]
\includegraphics[width=0.8\columnwidth]{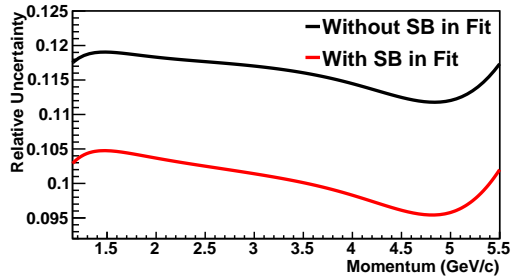}
\caption{Relative uncertainty on the double differential cross section as function of $K^+$ momentum (1.2$<P_{K}<$5.5~GeV/c) predicted by the F-S with and without including the SciBooNE production measurement.}%
\label{fig:unce_vs_mom}
\end{figure}

The relative uncertainties on the predicted double differential cross section by the F-S fit as function of $K^+$ angle and momentum decrease including the SciBooNE measurement as shown in Figures~\ref{fig:unce_vs_angle} and~\ref{fig:unce_vs_mom}.


The SciBooNE measurement confirms the validity of the F-S parametrization and including the SciBooNe measurement as an additional experimental data to the Feynman Scaling fit contributes in improving both the error uncertainty on the parametrization coefficients and in lowering the total uncertainty in the predicted $K^+$ production at 8.89~GeV/c proton momentum.

\subsection{SciBooNE $K^+$ Rate Measurement}


In addition to a measurement of $K^+$ production, the SciBooNE collaboration has also
published a measurement of the observed to MC predicted ratio for $K^+$ produced  neutrinos and
antineutrinos interacting in the SciBar detector. The results are summarized in Table~\ref{tab:nuance_fit_results_rate}.
The SciBooNE rate is the product of the $K^+$ production and neutrino cross section on carbon as explained in Ref.~\cite{Cheng:2011wq}. Since this result also includes the neutrino interaction cross section, it cannot be directly compared with the other experimental data presented in Table~\ref{Data Sets}. This constraint not only covers the neutrino flux from $K^+$ decay but also constrains the neutrino interaction cross section because the two targets are composed of similar material.
It should be noted that this analysis is a specific application to MiniBooNE and is not a general result. Nevertheless, the SciBooNE $K^+$ neutrino rate measurement can be directly applied to MiniBooNE analysis as a constraint on the electron and muon neutrinos from $K^+$ decay. Electron neutrinos from $K^+$ decays are one of the important background in the $\nu_\mu$ to $\nu_e$ oscillation search. Understanding this background will result in a reduction of the systematic uncertainty in the MiniBooNE oscillation analysis.

\begin{table}[htbp!]
  \begin{center}
    \vspace{0.5cm}
    \begin{ruledtabular}
    \begin{tabular}{cc|c|cc}
                                       &                                 		&                                		& Combined \\
                                       & $\nu$-mode             		& $\bar{\nu}$-mode   		& $\nu$+$\bar{\nu}$ mode \\ \hline
    $K^{+}$ Rate              & 0.94$\pm$0.05$\pm$0.11      & 0.54$\pm$0.09$\pm$0.30     & 0.88$\pm$0.04$\pm$0.10\\
    \end{tabular}
    \end{ruledtabular}
    \caption{$K^+$ rate measurement results relative to the MC beam prediction for the neutrino, antineutrino, and combined neutrino and anti-neutrino samples. Errors include statistical and systematic errors.}
  \label{tab:nuance_fit_results_rate}
  \end{center}
\end{table}

\begin{figure}[hptb!]
\includegraphics[width=0.8\columnwidth]{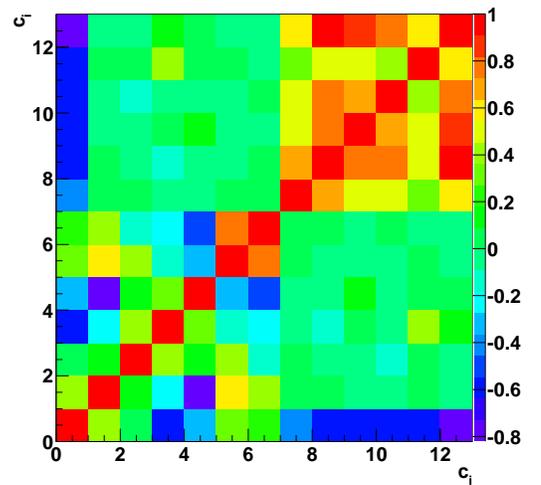}
\caption{Correlation between the fit parameters as in Figure~\ref{fig:Correlation_Matrix_after_SB_prod} but for the F-S fit results including the SciBooNE rate measurement.}%
\label{fig:Correlation_Matrix_after_SB_rate}%
\end{figure}

This SciBooNE $K^+$ rate measurement has been included in a version of the F-S fit and the best fit results for the parameters including the normalization for the data sets is reported in Table~\ref{tab:Fit_Results_after_SB_rate}.
The covariance matrix is reported in Table~\ref{tab:Covariance_Matrix_after_SB_rate} and correlation matrix is displayed in Figure~\ref{fig:Correlation_Matrix_after_SB_rate}.
Table~\ref{tab:Sweet_Spot_xsecs_after_SB_rate} lists the differential cross section values for kaon production at several kinematic points.

\begin{table}[htbp!]
\begin{center}
\begin{ruledtabular}
\begin{tabular}[c]{r|cc|r}
Scaling Fits & \multicolumn{2}{c|}{$1.2<P^{8.89}_{K}(GeV)<5.5$} & \\
& Value & Error &  \\\hline
c1 & 11.37 & 0.93 & \\
c2 & 0.87   & 0.13 & \\
c3 & 4.75   & 0.09 & \\
c4 & 1.51   & 0.06 & \\
c5 & 2.21   & 0.12 & \\
c6 & 2.17   & 0.43 & \\
c7 & 1.51   & 0.40 & Input Error\\\hline
Aleshin & 1.11 & 0.07 & 0.10\\
Allaby & 1.07 & 0.06 & 0.15\\
Dekkers & 0.87 & 0.06 & 0.20\\
Vorontsov & 0.54 & 0.04 & 5.00\\
Abbott & 0.78 & 0.07 & 0.15\\
Eichten & 1.03 & 0.06 & 0.15\\\hline
$\chi^{2}$/d.o.f. (no $f$) & 2.28 & (d.o.f. = 119) & \\
\end{tabular}
\end{ruledtabular}
\caption{Results for the F-S fits as in Figure~\ref{tab:Fit_Results_after_SB_prod} but for the F-S fit results including the SciBooNE rate measurement. d.o.f. indicates here degree of freedom and "no $f$" means no correction factor applied.}
\label{tab:Fit_Results_after_SB_rate}%
\end{center}
\end{table}

\begin{table*}[htbp!]
\begin{center}
\begin{ruledtabular}
\begin{tabular}[c]{r|ccccccc}
& c1 & c2 & c3 & c4 & c5 & c6 & c7\\\hline
c1  & 0.84        &   0.47E-01 & 0.39E-02 & -0.31E-01 & -0.36E-01 & 0.12   &   0.69E-01 \\
c2  & 0.47E-01 & 0.16E-01  & 0.14E-02 & -0.14E-02 & -0.13E-01 & 0.32E-01 &  0.22E-01\\
c3  & 0.40E-02 & 0.14E-02  & 0.73E-02  & 0.20E-02  & 0.19E-02 & 0.14E-01 & -0.33E-02 \\
c4  & -0.31E-01 & -0.14E-02 & 0.20E-02  & 0.34E-02 & 0.20E-02 & -0.38E-02 & -0.61E-02 \\
c5  & -0.36E-01 & -0.13E-01 & 0.19E-02  & 0.20E-02 & 0.15E-01 & -0.15E-01 & -0.24E-01 \\
c6  & 0.12         & 0.32E-01  & 0.14E-01 & -0.38E-02 & -0.15E-01 & 0.18  &    0.12   \\
c7  & 0.69E-01 & 0.22E-01  & -0.33E-02 & -0.61E-02 & -0.24E-01 & 0.12 &      0.16 \\
\end{tabular}
\end{ruledtabular}
\caption{Covariance matrix as in Table~\ref{tab:Covariance_Matrix_after_SB_prod} but for the F-S fit results including the SciBooNE rate measurement.}
\label{tab:Covariance_Matrix_after_SB_rate}
\end{center}
\end{table*}

\begin{table}[htbp!]
\begin{center}
\begin{ruledtabular}
\begin{tabular}[c]{l|cc|c}
& $P^{8.89}_{K}$ & $\theta_{K}$ & $\sigma_{K\text{ }prod}$ \\
& (GeV/c)  & (rad) & (mb)  \\\hline
$E_{\nu}=0.35$ GeV & 1.52 & 0.213 &   9.12 $\pm$ 0.62 (6.8\%) \\
$E_{\nu}=0.65$ GeV & 2.07 & 0.127 & 10.39 $\pm$ 0.62 (6.0\%)\\
$E_{\nu}=0.90$ GeV & 2.45 & 0.103 &  9.94 $\pm$ 0.58 (5.8\%) \\
Kaon Sweet Spot & 2.80 & 0.106 & 8.43 $\pm$ 0.49 (5.8\%) \\
HE $\nu_{\mu}$ Events & 4.30 & 0.055 & 4.60 $\pm$ 0.27 (5.8\%)\\
\end{tabular}
\end{ruledtabular}
\caption{Differential cross section values as in Table~\ref{tab:Sweet_Spot_xsecs_after_SB_prod} but for the F-S fit results including the SciBooNE rate measurement.}
\label{tab:Sweet_Spot_xsecs_after_SB_rate}%
\end{center}
\end{table}

In order to apply the SciBooNE constraint to the MiniBooNE neutrino event prediction, one needs to consider the $K^+$ kinematic regions that contribute to the two samples.

Figure~\ref{fig:kaon_that_produce_nues} shows that the kinematic region of $K^+$ mesons that produce
background $\nu_e$ events in MiniBooNE and Figure~\ref{fig:sciboone_nu_mode_E_vs_theta} shows the regions that contributes to the SciBooNE rate measurement.  While there is a large overlap between the SciBooNE and MiniBooNE regions, the MiniBooNE region extends to somewhat lower $K^+$ momenta. Using MC studies combined with the covariance matrix associated with F-S fit, we have quantified the increased uncertainty associated with extrapolating the SciBooNE measurement to the lower MiniBooNE region and found that the error on the constrained electron neutrino interaction rate should be increased by a factor of 1.5. This increases the uncertainty for the MiniBooNE electron neutrino event rate prediction from the measured SciBooNE uncertainty of 12\% (as reported in Table~\ref{tab:nuance_fit_results_rate}) to a total error of 18\%.  (The associated covariance matrix given in Table~\ref{tab:Covariance_Matrix_after_SB_rate} should also have all of the elements multiplied by $(1.5)^2=2.25$)  After applying the new SciBooNE constraint, the MiniBooNE prediction for electron neutrinos from $K^+$ decay is reduced by only 3\% but the uncertainty is reduced significantly by a factor of three from previous estimates because both the rate and cross section uncertainty is reduced \cite{MB_EDZ_Panic11}.


\section{Summary and Conclusions}

The F-S parameterization given in Eq.~\ref{Scaling_Function} has a theoretically motivated form that takes into account low beam momentum production thresholds from exclusive channels in contrast to many other models. For example, the S-W parameterization does not have the proper scaling properties or expected behavior for the $x_F<0$ regions. Also, extrapolations using data at much higher beam momentum appear to have difficulty describing lower momentum $K^{+}$ production measurements.

The F-S parameterization describes the $K^{+}$ production data well for beam momentum in the range of 8.89 to 24~GeV/c.
Fits involving different experimental data sets have been performed and show good agreement with the experimental data as shown in Figure~\ref{fig:fit_result_FS} where the data have been scaled by the normalization factors given in Table~\ref{Fit_Results_FS}. The normalization values (except for the Vorontsov data) are in good agreement within the 10\% to 20\% uncertainties quoted by the experiments.

The F-S fits including the full covariance matrix can be used to predict $K^{+}$ production for low beam momentum neutrino experiments such as the BNB at 8.89 GeV/c. The overall uncertainty from the fit is about 7\% and is consistent with the combination of the experiments with $\sim$15\% uncertainties. The fits also give the dependence on produced $K^{+}$ kinematics in angle and momentum, which is important for accurate neutrino flux predications using magnetic horn focusing devices.

A cross check of the F-S parameterization using neutrino data from the SciBooNE collaboration measurement reported in Ref.~\cite{Cheng:2011wq} confirms the accuracy of the model at low primary beam momenta and its validity as a better representation of $K^{+}$ production with respect to the S-W model. The F-S parameterization derived from the low energy kaon production experiments including this SciBooNE production constraint should therefore be a good representation of $K^{+}$ production for low energy neutrino beam simulations. We, therefore, suggest that the parameters shown in Table~\ref{tab:Fit_Results_after_SB_prod} be used along with the covariance given in Table~\ref{tab:Covariance_Matrix_after_SB_prod}.

\par We wish to acknowledge the MiniBooNE and SciBooNE Collaboration for the use of their neutrino simulation programs and the National Science Foundation for the support.



\bibliographystyle{apsrev}

\bibliography{references}

\begin{thebibliography}{23}
\expandafter\ifx\csname natexlab\endcsname\relax\def\natexlab#1{#1}\fi
\expandafter\ifx\csname bibnamefont\endcsname\relax
  \def\bibnamefont#1{#1}\fi
\expandafter\ifx\csname bibfnamefont\endcsname\relax
  \def\bibfnamefont#1{#1}\fi
\expandafter\ifx\csname citenamefont\endcsname\relax
  \def\citenamefont#1{#1}\fi
\expandafter\ifx\csname url\endcsname\relax
  \def\url#1{\texttt{#1}}\fi
\expandafter\ifx\csname urlprefix\endcsname\relax\def\urlprefix{URL }\fi
\providecommand{\bibinfo}[2]{#2}
\providecommand{\eprint}[2][]{\url{#2}}

\bibitem[{\citenamefont{Feynman}(1969)}]{FeynmanPRL}
\bibinfo{author}{\bibfnamefont{R.~P.} \bibnamefont{Feynman}},
  \bibinfo{journal}{Phys. Rev. Lett.} \textbf{\bibinfo{volume}{23}},
  \bibinfo{pages}{1415} (\bibinfo{year}{1969}).

\bibitem[{\citenamefont{Bonesini et~al.}(2001)\citenamefont{Bonesini,
  Marchionni, Pietropaolo, and Tabarelli~de Fatis}}]{Bonesini:2001iz}
\bibinfo{author}{\bibfnamefont{M.}~\bibnamefont{Bonesini}},
  \bibinfo{author}{\bibfnamefont{A.}~\bibnamefont{Marchionni}},
  \bibinfo{author}{\bibfnamefont{F.}~\bibnamefont{Pietropaolo}},
  \bibnamefont{and}
  \bibinfo{author}{\bibfnamefont{T.}~\bibnamefont{Tabarelli~de Fatis}},
  \bibinfo{journal}{Eur. Phys. J.} \textbf{\bibinfo{volume}{C20}},
  \bibinfo{pages}{13} (\bibinfo{year}{2001}), \eprint{hep-ph/0101163}.

\bibitem[{\citenamefont{Norbury}(2009)}]{Norbury:2009}
\bibinfo{author}{\bibfnamefont{J.~W.} \bibnamefont{Norbury}},
  \bibinfo{journal}{The Astrophysical Journal Supplement Series}
  \textbf{\bibinfo{volume}{182}}, \bibinfo{pages}{120} (\bibinfo{year}{2009}),
  \urlprefix\url{http://stacks.iop.org/0067-0049/182/i=1/a=120}.

\bibitem[{\citenamefont{Sanford and Wang}(1967)}]{SanWang}
\bibinfo{author}{\bibfnamefont{J.~R.} \bibnamefont{Sanford}} \bibnamefont{and}
  \bibinfo{author}{\bibfnamefont{C.~L.} \bibnamefont{Wang}}
  (\bibinfo{year}{1967}), \bibinfo{note}{bNL Internal Report number BNL 11479}.

\bibitem[{\citenamefont{Wang}(1970)}]{Wang:1970}
\bibinfo{author}{\bibfnamefont{C.~L.} \bibnamefont{Wang}},
  \bibinfo{journal}{Phys. Rev. Lett.} \textbf{\bibinfo{volume}{25}},
  \bibinfo{pages}{1068} (\bibinfo{year}{1970}),
  \urlprefix\url{http://link.aps.org/doi/10.1103/PhysRevLett.25.1068}.

\bibitem[{Sci()}]{SciBooNE}
\emph{\bibinfo{title}{Sciboone experiment}},
  \bibinfo{note}{http://www.sciboone.fnal.gov}.

\bibitem[{Min()}]{MiniBooNE}
\emph{\bibinfo{title}{Miniboone experiment}},
  \bibinfo{note}{http://www-boone.fnal.gov}.

\bibitem[{Mic()}]{MicroBooNE}
\emph{\bibinfo{title}{Microboone experiment}},
  \bibinfo{note}{http://www.microboone.fnal.gov}.

\bibitem[{\citenamefont{Kopp}(2007)}]{feynmanhiE}
\bibinfo{author}{\bibfnamefont{S.~E.} \bibnamefont{Kopp}},
  \bibinfo{journal}{Phys. Rept.} \textbf{\bibinfo{volume}{439}},
  \bibinfo{pages}{101} (\bibinfo{year}{2007}), \eprint{physics/0609129}.

\bibitem[{\citenamefont{Abbott et~al.}(1992)}]{abbott}
\bibinfo{author}{\bibfnamefont{T.}~\bibnamefont{Abbott}} \bibnamefont{et~al.}
  (\bibinfo{collaboration}{E-802}), \bibinfo{journal}{Phys. Rev.}
  \textbf{\bibinfo{volume}{D45}}, \bibinfo{pages}{3906} (\bibinfo{year}{1992}).

\bibitem[{\citenamefont{Aleshin et~al.}()\citenamefont{Aleshin, Drabkin, and
  Kolesnikov}}]{aleshin}
\bibinfo{author}{\bibfnamefont{Y.~D.} \bibnamefont{Aleshin}},
  \bibinfo{author}{\bibfnamefont{I.~A.} \bibnamefont{Drabkin}},
  \bibnamefont{and} \bibinfo{author}{\bibfnamefont{V.~V.}
  \bibnamefont{Kolesnikov}}, \emph{\bibinfo{title}{{Production of $K^{\pm}$
  Mesons from Be Targets at 62-Mrad at 9.5-GeV/c Incident Proton Momenta}}},
  \bibinfo{note}{iTEP-80-1977}.

\bibitem[{\citenamefont{Allaby et~al.}(1969)}]{allaby}
\bibinfo{author}{\bibfnamefont{J.~V.} \bibnamefont{Allaby}}
  \bibnamefont{et~al.}, \bibinfo{journal}{Phys. Lett.}
  \textbf{\bibinfo{volume}{B30}}, \bibinfo{pages}{549} (\bibinfo{year}{1969}).

\bibitem[{\citenamefont{Dekkers et~al.}(1965)}]{dekkers}
\bibinfo{author}{\bibfnamefont{D.}~\bibnamefont{Dekkers}} \bibnamefont{et~al.},
  \bibinfo{journal}{Phys. Rev.} \textbf{\bibinfo{volume}{137}},
  \bibinfo{pages}{B962} (\bibinfo{year}{1965}).

\bibitem[{\citenamefont{Eichten et~al.}(1972)}]{eichten}
\bibinfo{author}{\bibfnamefont{T.}~\bibnamefont{Eichten}} \bibnamefont{et~al.},
  \bibinfo{journal}{Nucl. Phys.} \textbf{\bibinfo{volume}{B44}},
  \bibinfo{pages}{333} (\bibinfo{year}{1972}).

\bibitem[{\citenamefont{Lundy et~al.}(1965)\citenamefont{Lundy, Novey,
  Yovanovitch, and Telegdi}}]{lundy}
\bibinfo{author}{\bibfnamefont{R.~A.} \bibnamefont{Lundy}},
  \bibinfo{author}{\bibfnamefont{T.~B.} \bibnamefont{Novey}},
  \bibinfo{author}{\bibfnamefont{D.~D.} \bibnamefont{Yovanovitch}},
  \bibnamefont{and} \bibinfo{author}{\bibfnamefont{V.~L.}
  \bibnamefont{Telegdi}}, \bibinfo{journal}{Phys. Rev. Lett.}
  \textbf{\bibinfo{volume}{14}}, \bibinfo{pages}{504} (\bibinfo{year}{1965}).

\bibitem[{\citenamefont{Marmer et~al.}(1969)}]{marmer}
\bibinfo{author}{\bibfnamefont{G.~J.} \bibnamefont{Marmer}}
  \bibnamefont{et~al.}, \bibinfo{journal}{Phys. Rev.}
  \textbf{\bibinfo{volume}{179}}, \bibinfo{pages}{1294} (\bibinfo{year}{1969}).

\bibitem[{\citenamefont{Pirou\'e and Smith}(1966)}]{piroue}
\bibinfo{author}{\bibfnamefont{P.~A.} \bibnamefont{Pirou\'e}} \bibnamefont{and}
  \bibinfo{author}{\bibfnamefont{A.~J.~S.} \bibnamefont{Smith}},
  \bibinfo{journal}{Phys. Rev.} \textbf{\bibinfo{volume}{148}},
  \bibinfo{pages}{1315} (\bibinfo{year}{1966}).

\bibitem[{\citenamefont{Vorontsov et~al.}()\citenamefont{Vorontsov, Safronov,
  Sibirtsev, Smirnov, and Trebukhovsky}}]{vorontsov}
\bibinfo{author}{\bibfnamefont{I.~A.} \bibnamefont{Vorontsov}},
  \bibinfo{author}{\bibfnamefont{G.~A.} \bibnamefont{Safronov}},
  \bibinfo{author}{\bibfnamefont{A.~A.} \bibnamefont{Sibirtsev}},
  \bibinfo{author}{\bibfnamefont{G.~N.} \bibnamefont{Smirnov}},
  \bibnamefont{and} \bibinfo{author}{\bibfnamefont{Y.~V.}
  \bibnamefont{Trebukhovsky}}, \emph{\bibinfo{title}{{A-Dependence of
  fragmentation of 9.2-GeV protons on nuclei. (in Russian)}}},
  \bibinfo{note}{iTEP-88-011}.

\bibitem[{\citenamefont{James and Roos}(1975)}]{minuit}
\bibinfo{author}{\bibfnamefont{F.}~\bibnamefont{James}} \bibnamefont{and}
  \bibinfo{author}{\bibfnamefont{M.}~\bibnamefont{Roos}},
  \bibinfo{journal}{Comput. Phys. Commun.} \textbf{\bibinfo{volume}{10}},
  \bibinfo{pages}{343} (\bibinfo{year}{1975}).

\bibitem[{\citenamefont{Aguilar-Arevalo et~al.}(2009)}]{AguilarArevalo:2008yp}
\bibinfo{author}{\bibfnamefont{A.~A.} \bibnamefont{Aguilar-Arevalo}}
  \bibnamefont{et~al.} (\bibinfo{collaboration}{MiniBooNE}),
  \bibinfo{journal}{Phys. Rev.} \textbf{\bibinfo{volume}{D79}},
  \bibinfo{pages}{072002} (\bibinfo{year}{2009}), \eprint{0806.1449}.

\bibitem[{\citenamefont{Taylor et~al.}(1976)\citenamefont{Taylor, Carey,
  Johnson, Kammerud, Ritchie, Roberts, Sauer, Shafer, Theriot, and
  Walker}}]{taylor:1976}
\bibinfo{author}{\bibfnamefont{F.~E.} \bibnamefont{Taylor}},
  \bibinfo{author}{\bibfnamefont{D.~C.} \bibnamefont{Carey}},
  \bibinfo{author}{\bibfnamefont{J.~R.} \bibnamefont{Johnson}},
  \bibinfo{author}{\bibfnamefont{R.}~\bibnamefont{Kammerud}},
  \bibinfo{author}{\bibfnamefont{D.~J.} \bibnamefont{Ritchie}},
  \bibinfo{author}{\bibfnamefont{A.}~\bibnamefont{Roberts}},
  \bibinfo{author}{\bibfnamefont{J.~R.} \bibnamefont{Sauer}},
  \bibinfo{author}{\bibfnamefont{R.}~\bibnamefont{Shafer}},
  \bibinfo{author}{\bibfnamefont{D.}~\bibnamefont{Theriot}}, \bibnamefont{and}
  \bibinfo{author}{\bibfnamefont{J.~K.} \bibnamefont{Walker}},
  \bibinfo{journal}{Phys. Rev. D} \textbf{\bibinfo{volume}{14}},
  \bibinfo{pages}{1217} (\bibinfo{year}{1976}),
  \urlprefix\url{http://link.aps.org/doi/10.1103/PhysRevD.14.1217}.

\bibitem[{\citenamefont{Cheng et~al.}(2011)}]{Cheng:2011wq}
\bibinfo{author}{\bibfnamefont{G.}~\bibnamefont{Cheng}} \bibnamefont{et~al.}
  (\bibinfo{collaboration}{SciBooNE Collaboration}),
  \bibinfo{journal}{Phys.Rev.} \textbf{\bibinfo{volume}{D84}},
  \bibinfo{pages}{012009} (\bibinfo{year}{2011}), \eprint{1105.2871}.

\bibitem[{\citenamefont{Zimmerman}(2011)}]{MB_EDZ_Panic11}
\bibinfo{author}{\bibfnamefont{E.}~\bibnamefont{Zimmerman}}, in
  \emph{\bibinfo{booktitle}{Proceeding of The 19th Particles and Nuclei
  International Conference (PANIC11)}} (\bibinfo{year}{2011}).

\end{thebibliography}

\end{document}